# Sky Surveys

S. G. Djorgovski, A.A. Mahabal, A.J. Drake, M.J. Graham, and C. Donalek

California Institute of Technology
Pasadena, CA 91125, USA
george@astro.caltech.edu

**Abstract:**

Sky surveys represent a fundamental data basis for astronomy. We use them to map in a systematic way the universe and its constituents, and to discover new types of objects or phenomena. We review the subject, with an emphasis on the wide-field, imaging surveys, placing them in a broader scientific and historical context. Surveys are now the largest data generators in astronomy, propelled by the advances in information and computation technology, and have transformed the ways in which astronomy is done. This trend is bound to continue, especially with the new generation of synoptic sky surveys that cover wide areas of the sky repeatedly, and open a new time domain of discovery. We describe the variety and the general properties of surveys, illustrated by a number of examples, the ways in which they may be quantified and compared, and offer some figures of merit that can be used to compare their scientific discovery potential. Surveys enable a very wide range of science, and that is perhaps their key unifying characteristic. As new domains of the observable parameter space open up thanks to the advances in technology, surveys are often the initial step in their exploration. Some science can be done with the survey data alone (or a combination of data from different surveys), and some requires a targeted follow-up of potentially interesting sources selected from surveys. Surveys can be used to generate large, statistical samples of objects that can be studied as populations, or as tracers of larger structures to which they belong. They can be also used to discover or generate samples of rare or unusual objects, and may lead to discoveries of some previously unknown types. We discuss a general framework of parameter spaces that can be used for an assessment and comparison of different surveys, and the strategies for their scientific exploration. As we are moving into the Petascale regime and beyond, an effective processing and scientific exploitation of such large data sets and data streams poses many challenges, some of which are specific to any given survey, and some of which may be addressed in the framework of Virtual Observatory and Astroinformatics. The exponential growth of data volumes and complexity makes a broader application of data mining and knowledge discovery technologies critical in order to take a full advantage of this wealth of information. Finally, we discuss some outstanding challenges and prospects for the future.

---

**Index Terms**:

Sky surveys; synoptic surveys; wide-field surveys; deep surveys; catalogs; archives; technology; virtual observatory (VO); astroinformatics; data processing pipelines; digital imaging; astronomical photography; observable parameter space (OPS); measurement parameter space (MPS); physical parameter space (PPS); time domain; multi-wavelength astronomy; data mining; classification; systematic exploration; figures of merit for sky surveys; statistical studies; software; pipelines; history of astronomy.

**Keywords**:

Sec. 1: Sky surveys; synoptic surveys; wide-field surveys; deep surveys; catalogs; archives; virtual observatory (VO); digital imaging; astronomical photography; multi-wavelength astronomy.

Sec. 2: Sky surveys; history of astronomy; astronomical photography; space-based astronomy.

Sec. 3: Technology; observable parameter space (OPS); measurement parameter space (MPS); physical parameter space (PPS); time domain; multi-wavelength astronomy; data mining; classification; systematic exploration; statistical studies.

Sec. 4: Sky surveys; synoptic surveys; wide-field surveys; deep surveys; catalogs; archives; multi-wavelength astronomy; supernova surveys; asteroid surveys; microlensing surveys; time domain; figures of merit for sky surveys.

Sec. 5: Software; technology; data processing pipelines; archives; virtual observatory (VO); astroinformatics; data mining; classification; time domain.

Sec. 6:

Entire Chapter: Same as the Index Terms.

**Cross-References:**

Astronomical Software
IR astronomy
Large Scale Structure of the Universe
Overview of Instrumentation and Detectors
Robotic and Survey Telescopes
Silicon-based Imaging Sensors
Statistical Methods for Astronomy
Virtual Observatories, Data Mining, and Astroinformatics

**List of Selected Abbreviations**:
AAVSO = American Association of Variable Star Observers, http://www.aavso.org/
CCD = Charge Coupled Device
CfA = Harvard-Smithsonian Center for Astrophysics, http://cfa.harvard.edu
CFHT = Canada-France-Hawaii Telescope, http://www.cfht.hawaii.edu
CMBR = Cosmic Microwave Background Radiation
EB = Exabyte ($10^{18}$ bytes)
ESO = European Southern Observatory, http://eso.org
FITS = Flexible Image Transport System, http://heasarc.nasa.gov/docs/heasarc/fits.html
FoM = Figure of Merit
FOV = Field of view
FWHM = Full Width at Half Maximum
GB = Gigabyte ($10^9$ bytes)
HST = Hubble Space Telescope, http://www.stsci.edu/hst
ICT = Information and Computing Technology
LSS = Large-Scale Structure
MB = Megabyte ($10^6$ bytes)
NOAO = National Optical Astronomy Observatory, http://noao.edu
NRAO = National Radio Astronomy Observatory, http://nrao.edu
MJD =  Modified Julian Date
MPS = Measurement Parameter Space
OPS = Observable Parameter Space
PB = Petabyte ($10^{15}$ bytes)
PPS = Physical Parameter Space
SETI = Search for Extraterrestrial Intelligence, http://www.seti.org
SN = Supernova
TB = Terabyte ($10^{12}$ bytes)
USNO = United States Naval Observatory, http://www.usno.navy.mil/
VLA = NRAO Very Large Array, http://www.vla.nrao.edu
VO = Virtual Observatory, http://www.ivoa.net
WWT = WorldWide Telescope, http://www.worldwidetelescope.org

Additional abbreviations for the various sky surveys and catalogs are listed in the Appendix.

# 1. Introduction

## *1.1 Definitions and Caveats*

Sky surveys are at the historical core of astronomy. Charting and monitoring the sky gave rise to our science, and today large digital sky surveys are transforming the ways astronomy is done. In this chapter we review some of the general issues related to the strategic goals, planning, and execution of modern sky surveys, and describe some of the currently popular ones, at least as of this writing (late 2011). This is a rapidly evolving field, and the reader should consult the usual sources of information about the more recent work.

Some caveats are in order: The very term "sky surveys" is perhaps too broad and loosely used, encompassing a very wide range of the types of studies and methods. Thus, we focus here largely on the wide-field, panoramic sky surveys, as opposed, e.g., to specific studies of deep fields, or to heavily specialized surveys of particular objects, types of measurements, etc. We also have a bias towards the visible regime, reflecting, at least partly, the authors' expertise, but also a deeper structure of astronomy: most of the science often requires a presence of a visible counterpart, regardless of the wavelength coverage of the original detection. We also focus mainly on the imaging surveys, with a nod to the spectroscopic ones. However, many general features of surveys in terms of the methods, challenges, strategies, and so on, are generally applicable across the wavelengths and types of observations. There are often no sharp boundaries between different kinds of surveys, and the divisions can be somewhat arbitrary. Finally, while we outline very briefly the kinds of science that is done with sky surveys, we do not go into any depth for any particular kind of studies or objects, as those are covered elsewhere in these volumes.

It is tempting to offer a working definition of a survey in this context. By a *wide-field survey*, we mean a large data set obtained over areas of the sky that may be at least of the order of $\sim 1\%$ of the entire sky (admittedly an arbitrary choice), that can support a variety of scientific studies, even if the survey is devised with a very specific scientific goal in mind. However, there is often a balance between the depth and the area coverage. A *deep survey* (e.g., studies of various deep fields) may cover only a small area and contain a relatively modest number of sources by the standards of wide-field surveys, and yet it can represent a survey in its own right, feeding a multitude of scientific studies. A wide-field coverage by itself is not a defining characteristic: for example, all-sky studies of the CMBR may be better characterized as focused experiments, rather than as surveys, although that boundary is also getting fuzzier.

We also understand that "a large data set" is a very relative and rapidly changing concept, since data rates increase exponentially, following Moore's law, so perhaps one should always bear in mind a conditional qualifier "at that time". Some types of surveys are better characterized by a "large" number of sources (the same time-dependent conditional applies), which is also a heavily wavelength-dependent measure; for example, nowadays a thousand is still a large number of γ-ray sources, but a trivial number of visible ones.

Perhaps the one unifying characteristic is that surveys tend to support a broad variety of studies, many of which haven't been thought of by the survey's originators. Another unifying characteristic is the exploratory nature of surveys, which we address in more detail in Sec. 3 below. Both approaches can improve our knowledge of a particular scientific domain, and can lead to surprising new discoveries.

The meaning of the word "survey" in the astronomical context has also changed over the years. It used to refer to what we would now call a sky atlas (initially hand-drawn sky charts, and later photographic images), whereas catalogs of sources in them were more of a subsidiary or derived data product. Nowadays the word largely denotes catalogs of sources and their properties (positions, fluxes, morphology, etc.), with the original images provided almost as a subsidiary information, but with an understanding that sometimes they need to be reprocessed for a particular purpose. Also, as the complexity of data increased, we see a growing emphasis on carefully documented metadata ("data about the data") that are essential for the understanding of the coverage, quality, and limitations of the primary survey data.

*1.2 The Types and Goals of Sky Surveys*

We may classify surveys in regard to their scientific motivation and strategy, their wavelength regime, ground-based vs. space-based, the type of observations (e.g., imaging, spectroscopy, polarimetry, etc.), their area coverage and depth, their temporal character (one-time vs. multi-epoch), as panoramic (covering a given area of the sky with all sources therein) or targeted (observing a defined list of sources), and can have any combination of these characteristics. For example, radio surveys generally produce data cubes, with two spatial and one frequency dimension, and are thus both imaging and spectroscopic, and often including the polarization as well. X-ray and γ-ray images generally also provide some energy resolution. Slitless spectroscopy surveys (images taken through an objective prism, grating, or a grism) provide wavelength-dispersed images of individual sources. Surveys can be also distinguished by their angular, temporal, or energy resolution.

Surveys may be scientifically motivated by a census of particular type of sources, e.g., stars, galaxies, or quasars, that may be used for statistical studies such as the Galactic structure or the Large-Scale Structure (LSS) in the universe. They may be aimed to discover significant numbers of a particular type of objects, often relatively rare ones, for the follow-up studies, e.g., Supernovae (SNe), high-redshift galaxies or quasars, brown dwarfs, etc. When a new domain of an observable parameter space opens up, e.g., a previously unexplored wavelength regime, it usually starts with a panoramic survey, to see what kinds of objects or phenomena populate it.

Therein lies perhaps the key scientific distinction between surveys and the traditional, targeted astronomical observations: surveys aim to map and characterize the astrophysical contents of the sky or of the populations of objects of particular kinds in a systematic manner, whereas the traditional observations focus on detailed properties of individual sources or relatively small numbers of them. Surveys are often the ways to find such targets for detailed studies.

The first type of survey science – use of large, statistical samples of objects of some kind (stars, galaxies, etc.) as probes of some collective properties (e.g., Galactic structure, or LSS) – may be done with the survey data alone, or may be supplemented by additional data from other sources. The other two types of survey science – as a discovery mechanism for rare, unusual, or new types of objects or phenomena, of as a pure initial exploration of some new domain of the observable parameter space – require targeted follow-up observations. Thus surveys become a backbone of much of astronomical research today, forming a fundamental data infrastructure of astronomy. This may make them seem less glamorous than the successful targeted observations that may be enabled by surveys, but in does not diminish their scientific value.

Imaging surveys are commonly transformed into catalogs of detected sources and their properties, but in some cases images themselves represent a significant scientific resource, e.g., if they contain extended structures of diverse morphologies; for example, images of star-forming regions, or stellar bubbles and SN remnants in H$\alpha$ images.

The process of detection and characterization of discrete sources in imaging surveys involves many challenges and inevitably introduces biases, since these processes always assume that the sources have certain characteristics in terms of a spatial extent, morphology, and so on. We discuss these issues further in Sec. 5.

Like most astronomical observations, surveys are often enabled by new technologies, and push them to their limits. Improved detector and telescope technologies can open new wavelength regimes, or more sensitivity or resolution, thus providing some qualitatively new view of the sky. A more recent phenomenon is that information and computation technologies (ICT) dramatically increased our ability to gather and process large quantities of data, and that quantitative change has led to some interesting qualitative changes in the ways we study the universe.

A direct manifestation of this is the advent of large synoptic sky surveys, that cover large areas of the sky repeatedly and often, thus opening the time domain as new arena for exploration and discovery. They are sometimes described as a transition from a panoramic cosmic photography to a panoramic cosmic cinematography. We describe some examples below.

Spectroscopic surveys, other than the data cubes generated in radio astronomy, typically target lists of objects selected from imaging surveys. In case of extragalactic surveys, the primary goal is typically to obtain redshifts, as well as to determine some physical properties of the targets, e.g., star formation rates, or presence and classification of active galactic nuclei (AGN), if any. If the targets are observed with long slit or multi-slit mask spectrographs, or integral field units (IFU), information can be obtained about the kinematics of resolved structures in galaxies, typically emission-line gas. In case of Galactic survey, the goals are typically to measure radial velocities, and sometimes also the chemical abundances of stars.

Spectroscopic surveys depend critically on the quality of the input catalogs from which the targets are selected, inheriting any biases that may be present. Their observing strategies in terms of the depth, source density, spectroscopic resolution, etc., are determined by the scientific goals. Since spectroscopy is far more expensive than imaging in terms of the observing time, some redshift surveys have adopted a sparse-sampling strategy, e.g., by observing every $N^{th}$ (where N = 2, or 10, or…) source in a sorted list of targets, thus covering a larger area, but with a corresponding loss of information.

Our observations of the sky are no longer confined to the electromagnetic window. Increasingly, sky is being monitored in high-energy cosmic rays (Kotera & Olinto 2011), neutrinos (Halzen & Klein 2010), and even gravitational waves (Centrella 2010). So far, these information channels have been characterized by a paucity of identified sources, largely due to the lack of a directional accuracy, with the exceptions of the Sun and SN 1987A in neutrinos, but they will likely play a significant role in the future.

Finally, as numerical simulations become ever larger and more complex, and theory is expressed as data (the output of simulations), we may start to see surveys of simulations, as means of characterizing and quantifying them. These surveys of theoretical universes would have to be compared to the measurements obtained in the surveys of the actual universe. New knowledge often arises as theories are confronted with data, and in the survey regime, we will be doing that on a large scale.

*1.3 The Data Explosion*

In 1990's, astronomy transitioned from a relatively data-poor science to an immensely data-rich one, and the principal agent of change were large digital sky surveys. They, in turn, were enabled by the rapid advances in ICT. Sky surveys became the dominant data sources in astronomy, and this trend continues (Brunner et al. 2001c). The data volume in astronomy doubles at Moore's law pace, every year to a year and a half (Szalay & Gray 2001, Gray & Szalay 2006), reflecting the growth of the technology that produces the data. (Obviously, the sheer size of data sets by itself does not imply a large scientific value; for example, very deep images from space-based observatories may have a modest size in bits, but an immense scientific value.)

In the past, surveys and their derived catalogs could be published as printed papers or small sets of volumes that can be looked up "by hand" (this is still true in some regimes, e.g., the γ-ray astronomy, or other nascent fields). But as the data volumes entered the Terascale regime in the 1990's, and the catalogs of sources started containing millions of objects, there was an inevitable transition to a purely electronic publication and dissemination, e.g., in the form of the web-accessible archives, that also provide access to the necessary metadata and other documentation. Databases, data mining, web services, and other computational tools and techniques, became a standard part of astronomy's tool chest, although the community is still gradually gaining their familiarity with them. This is an aspect of an inevitable culture change, as we enter the era of a data-rich, data-intensive science.

The growth of data quantity, coupled with an improved data homogeneity, enabled a new generation of statistical or population studies: with samples of hundreds of millions of sources, the Poissonian errors were no longer important, and one could look for subtle effects simply not accessible with the more limited data sets. Equally important was the growth of data quality and data complexity. The increased information content of the modern sky surveys enabled a profitable data mining: the data could be used for a much broader variety of studies than it was possible in the past.

For these reasons, survey-enabled astronomy became both popular and respectable. But it was obvious that data fusion across different surveys (e.g., over different wavelengths) has an even higher scientific potential, as it can reveal knowledge that is present in the combined data, but cannot be recognized in any individual data set, no matter how large. Historical examples from multi-wavelength cross-correlations abound, e.g., the discoveries of quasars, ultraluminous starbursts, interpretation of γ-ray bursts, etc. The new, data-rich astronomy promised to open this discovery arena wholesale.

There are many non-trivial challenges posed by the handling of large, complex data sets, and knowledge discovery in them: how to process, and calibrate the raw data; how to store, combine, and access them using modern computing hardware and networks; and how to visualize, explore and analyses these great data sets quickly and efficiently. This is a rapidly developing field, increasingly entails collaborative efforts between astronomers and computer scientists.

The rise of data centers was the response to dealing with *individual* large data sets, surveys, or data collections. However, their fusion and the scientific synthesis required more than just their interoperability. This prompted the rise of the *Virtual Observatory* (VO) concept, as a general, distributed research environment for astronomy with large and complex data sets (Brunner et al. 2001a, Hanisch 2001, 2010, Djorgovski & Williams 2005). Today, sky surveys are naturally included in an evolving world-wide ecosystem of astronomical data resources and services. The reader is directed to the VO-related websites or their future equivalents, for an up to date description of the data assets and services, and access to them.

Astronomy was not alone in facing the challenges and the opportunities of an exponential data growth. Virtual scientific organizations with similar mandates emerged in many other fields, and continue to do so. This entire arena of a computationally-enabled, data-driven science is sometimes referred to as Cyber-Infrastructure, or e-Science, unified by the common challenges and new scientific methodologies (Atkins et al. 2003, Hey & Trefethen 2003, 2005, Djorgovski 2005, Hey et al. 2009, Bell et al. 2009). Nowadays we also see the blossoming of "science informatics", e.g., Astroinformatics (by analogy with its bio-, geo-, etc., counterparts). These are broader concepts of scientific and methodological environments and communities of interest, that seek to develop and apply new tools for the data-rich science in the $21^{st}$ century.

## 2. A (Very) Brief History of Sky Surveys

We cannot do justice to this subject in the limited space here, and we just mention some of the more important highlights, in order to put the subject in a historical context. The interested reader may wish to start with a number of relevant, excellent Wikipedia articles and other Web resources for additional information and references.

Historically, surveying of the sky started with a naked eye (that may be looking through a telescope), and we could consider Charles *Messier*'s catalog from the middle of the $18^{th}$ century as a forerunner of the grander things to come. In the pre-photography era, the most notable sky surveying was done by the Herschel family (brother William, sister Caroline, and son John), starting in the late $18^{th}$ century. Among the many notable achievements, the Herschels also introduced stellar statistics – counting of stars per unit area, that can be used to learn more about our place in the universe. Their work was continued by many others, leading to the publication of the first modern catalogs in the late $19^{th}$ centrury, e.g., the still-used *New General Catalogue* (NGC) and *Index Catalogue* (IC) by John Dreyer (Dreyer 1888, 1895).

Visually compiled star catalogs culminated in the mid/late $19^{th}$ century, including the *Bonner Durchmusterung* (BD) in the North, its Southern equivalent, the *Cordoba Durchmusterung* (CD), that contained nearly 900,000 stars.

The field was transformed by the advent of a new technology – photography. Surveys that may be recognized as such in the modern sense of the term started with the first photographic efforts that covered systematically large areas of the sky at the end of the $19^{th}$ century. Perhaps the most notable of those is the *Harvard Plate Collection*, that spans over a century of sky coverage, and that is currently being scientifically rejuvenated through digitization by the *Digital Access to Sky Century at Harvard* project (DASCH; http://hea-www.harvard.edu/DASCH; Grindlay et al. 2009). A roughly contemporaneous, international effort, *Carte du Ciel*, led to the production of the *Astrographic Catalogue* (AC), reaching to $m \sim 11$ mag, that served as the basis of the more modern astrometric catalogs.

Another important innovation introduced at the Harvard College Observatory at the turn of the $19^{th}$ century was systematic monitoring of selected areas on the sky, a precursor of the modern synoptic sky surveys. Photographic monitoring of the Magellanic Clouds enabled Henrietta Leavitt (Leavitt & Pickering 1912) to discover the period-luminosity relations for Cepheids, thus laying the groundwork for the cosmological distance scale and the discovery of the expanding universe by Edwin Hubble and others in the 1920's.

In the early decades of the 20$^{th}$ century, Edward Pickering, Annie Jump Cannon, and their collaborators at Harvard produced the *Henry Draper Catalogue* (HD), named after the donor, that was eventually extended to ~ 360,000 stars, giving spectral types based on objective prism plates. Around the same time, Harlow Shapley and collaborators catalogued for the first time tens of thousands of galaxies in the Southern sky, and noted the first signs of the large-scale structure in the universe.

Around the same time, roughly the first third of the 20$^{th}$ century, the Dutch school (Jakobus Kapteyn, Pieter van Rhijn, Jan Oort, Bart Bok, and their students and collaborators) started systematic mapping of the Milky Way using star counts, and laid foundations for the modern studies of Galactic structure. Kapteyn also introduced *Selected Areas* (SA), a strategically chosen set of directions where star counts can yield information about the Galactic structure, without the need to survey the entire sky. In 1966, the heavily used *Smithsonian Astrophysical Observatory Catalog* was published, that contained positions, proper motions, magnitudes, and (usually) spectral types for over 250,000 stars.

A key figure, emerging in the 1930's was Fritz Zwicky, who, among many other ideas and discoveries, pioneered systematic sky surveys in the ways that shaped much of the subsequent work. He built the first telescope on Mt. Palomar, the 18-inch Schmidt, then a novel design for a wide-field instrument. With Walter Baade, he used it to search for Supernovae (a term he and Baade introduced), leading to the follow-up studies and the physical understanding of this phenomenon (Baade & Zwicky 1927). Zwicky's systematic, panoramic mapping of the sky led to many other discoveries, including novel types of compact dwarf galaxies, and more evidence for the LSS. In 1960's, he and his collaborators E. Herzog, P. Wild, C. Kowal, and M. Karpowitz published the *Catalogue of Galaxies and of Clusters of Galaxies* (CGCG; 1961-1968), that served as an input for many redshift surveys and other studies in the late 20$^{th}$ century. All this established a concept of using wide-field surveys to discover rare or interesting objects to be followed up by larger instruments.

The potential of Schmidt telescopes as sky mapping engines was noted, and a 48-inch Schmidt was built at Palomar Mountain, largely as a means of finding lots of good targets for the newly built 200-inch, for a while the largest telescope in the world. A major milestone was the first *Palomar Observatory Sky Survey* (POSS-I), conducted from 1949 to 1958, and spearheaded mainly by Edwin Hubble, Milton Humason, Walter Baade, Ira Bowen, and Rudolph Minkowski, and was funded mainly by the National Geographic Society (Minkowski & Abell 1963).

POSS-I mapped about 2/3 of the entire sky, observable from Palomar Mountain, initially from the North Celestial Pole down to Dec ~ –30°, and later extended to Dec ~ –42° (the Whiteoak extension). The survey used 14-inch wide photographic plates, covering roughly 6.5°× 6.5° fields of view (FOV) each, but with a useful, unvignetted FOV of ~ 6°× 6°, with some overlaps, with a total of 936 fields in each of the two bandpasses, one using the blue-sensitive Kodak 103a-O plates, and one using the red sensitive Kodak 103a-F emulsion. Its limiting magnitudes vary across the survey, but are generally close to $m_{lim}$ ~ 21 mag. Reproduced as glass copies of the original plates and as paper prints, POSS-I served as a fundamental resource, effectively a roadmap for astronomy, for several decades, and in its digital form it is still used today.

Its cataloguing was initially done by eye, and some notable examples include the *Uppsala General Catalogue* (UGC) of ~ 13,000 galaxies with apparent angular diameters > 1 arcmin (Nilson 1973) and *Morphological Catalog of Galaxies* (MCG) containing ~ 30,000 galaxies down to $m$ ~ 15 mag (Vorontsov-Velyaminov & Arkhipova 1974); Abell's (1958) catalog of ~ 2,700 clusters of galaxies; and many others.

The *Second Palomar Observatory Sky Survey* (POSS-II), conducted about 4 decades later, was the last of the major photographic sky surveys (Reid et al. 1991). Using an improved telescope optics and improved photographic emulsions, it covered the entire Northern sky with ~ 900 partly overlapping 6.5° fields spaced by 5°, in 3 bandpasses, corresponding to Kodak IIIa-J (blue), IIIa-F (red) and IV-N (far red) emulsions.

A number of other surveys have been conducted at the Palomar 48-inch Schmidt (renamed to *Samuel Oschin Telescope*, in honor of the eponymous benefactor) in the intervening years, including Willem Luyten's measurements of stellar proper motions, the original *Hubble Space Telescope Guide Star Catalog* (GSC), and a few others.

Southern sky equivalents of the POSS surveys in terms of the coverage, depth, and distribution were conducted at the European Southern Observatory's 1.0-m Schmidt telescope, and the 1.2-m UK Schmidt at the Anglo-Australian Observatory in the 1970's and 1990's, jointly called the *ESO/SERC Southern Sky Survey*. Andris Lauberts (1982) produced a Southern sky equivalent of the UGC catalog.

Both POSS and ESO/SERC surveys have been digitized independently by several groups in the 1990's. Scans produced at the Space Telescope Science Institute were used to produce both the second generation HST Guide Star Catalog (GSC-2; Lasker et al. 2008), and the *Digital Palomar Observatory Sky Survey* (DPOSS; Djorgovski et al. 1997a, 1999); the images are distributed through several *Digitized Sky Survey* (DSS) servers world-wide. The US Naval Observatory produced the astrometric *USNO-A* and *USNO-B* catalogs that also include proper motions (Monet et al. 2003). These digital versions of photographic sky surveys are described in more detail below. The surveys were also scanned by the *Automated Plate Measuring* facility (http://www.ast.cam.ac.uk/~mike/casu; APM) in Cambridge, the S*uperCOSMOS* group (http://www-wfau.roe.ac.uk/sss; Hambly et al. 2001), and the *Automated Plate Scanner* group (APS; http://aps.umn.edu; Cabanela et al. 2003). Scans of the ESO/SERC Southern Sky survey plates resulted in the *APM Galaxy Survey* (Maddox et al. 1990ab). DSS scans and other surveys can be also accessed through *SkyView* (http://skyview.gsfc.nasa.gov; McGlynn et al. 1997).

The middle of the 20[th] century also saw an appearance of a plethora of sky surveys at other wavelengths, as the new regimes opened up (IR, radio, X-ray, γ-ray, etc.). Enabled by the new technologies, e.g., electronics, access to space, computers, etc., these new, panchromatic views of the sky led to the discoveries of many previously unknown types of objects and phenomena, e.g., quasars, pulsars, various X-ray sources, protostars, or γ-ray bursts, to name but a few.

A good account of the history of radio astronomy is by Sullivan (2009). Following the pioneering work by Karl Jansky and Grote Reber, radio astronomy started to blossom in the 1950's, fueled in part by the surplus radar equipment from the World War II. A key new development was the technique of aperture synthesis and radio interferometers, pioneered by Martin Ryle, Anthony Hewish, and collaborators. This led to the first catalogs of radio sources, the most enduring of which were the Third Cambridge (3C) and Fourth Cambridge (4C) catalogs, followed by the plethora of others. Optical identifications of radio sources by Walter Baade, Rudolph Minkowski, Maarten Schmidt, and many others in the 1950's and 1960's opened whole new areas of research that are still thriving today. Among the very many surveys for AGN and/or objects with a UV excess, we can mention the classic Palomar Green (PG; Green 1976) survey, and an extensive body of work by Benjamin Markarian at Byurakan Observatory in Armenia, from 1960's to 1980's.

While the infrared (IR) light was discovered already by William Herschel circa 1800, the IR astronomy started in earnest with the advent of first efficient IR detectors in the 1960's. A pioneering 2 μm IR sky survey (TMASS) was done by Gerry Neugebauer and Robert Leighton (1969), resulting in a catalog of ~ 5,600 sources. Its modern successor some 3 decades later, the *Two Micron All-Sky Survey* (2MASS), catalogued over 300 million in three bandpasses (*JHK*); it is described in more detail below. Deeper surveys followed, notably the *United Kingdom Infrared Telescope (UKIRT) Infrared Deep Sky Survey* (UKIDSS), and more are forthcoming.

A wholesale exploration of the mid/far IR regime required a space-borne platform, and the *Infrared Astronomy Satellite* (IRAS), launched in 1983, opened a huge new area of research, continued by a number of space IR missions since. Some of the milestones include the DIRBE and FIRAS experiments on the *Cosmic Background Explorer satellite* (COBE), launched in 1989, and, more recently the *Wide-field Infrared Survey Explorer* (WISE), launched in 2009.

Astronomy at higher energies, i.e., UV beyond the atmospheric cutoff limit, X-rays, and γ-rays required access to space, that was a beneficent product of the space race and the cold war, starting shortly after the World War II.

The UV (i.e., with λ < 320 nm or so) astronomy from space started with a number of targeted missions starting from the 1960's, the first, shallow all-sky survey was done with the TD-1A satellite in 1972, followed by the *Extreme UV Explorer* (EUVE, 1992-2001). The next major milestone was the *Galaxy Evolution Explorer* (GALEX, 2003-2011), that surveyed most of the sky in two broad bands, down to the UV equivalent of the optical POSS surveys, and a much smaller area about 2 – 3 mag deeper.

Following the birth of the X-ray astronomy with rocket-borne experiments in the 1960's, the first X-ray all-sky surveys started with the SAS-A *Uhuru* (1970-1973) satellite, HEAO-1 mission (1977-1979), and HEAO-2 *Einstein* (1978-1981), followed by many others, including *Rosat* (1990-1999). Many other missions over the past 3 decades followed the survey ones with pointed observations of selected targets. They led to the discoveries of many new aspects of the previously known types of sources, e.g., accreting binaries, active galactic nuclei, clusters of galaxies, etc., and provided key new insights into their nature.

Gamma-ray astronomy is still mostly done with space-borne instruments that cover most or all of the sky. This is largely because γ-rays are hard to focus and shield from, so the instruments tend to look at all directions at once. While the first cosmic γ-ray emission was detected in 1961 by the Explorer-11 satellite, with a total of about 100 photons collected, γ-rays surveys really started with the SAS-2 (1972) and the COS-B (1975–1982) satellites. A major milestone was the *Compton Gamma-Ray Observatory* (CGRO, 1991-2000), followed by the *Fermi Gamma-Ray Space Telescope* (FGST), launched in 2008. These and other missions uncovered many important phenomena in the high-energy universe, but perhaps the most spectacular are the cosmic γ-ray bursts (GRBs), discovered in 1967 by the *Vela* satellites, and finally understood in 1997 thanks to the *BeppoSAX* mission.

Descriptions of all of these fascinating discoveries are beyond the scope of this chapter, and can be found elsewhere in these volumes.

In the 1990's, the era of fully digital sky surveys began in earnest. Aside from the digitized versions of the photographic sky surveys, a major milestone was the *Sloan Digital Sky Survey* (SDSS), which really helped change the culture of astronomy. Another key contribution at that time was the *Two Micron All-Sky Survey* (2MASS). Radio astronomy contributed the *New VLA Sky Survey* (NVSS) and the *Faint Images of Radio Sky at Twenty centimeters* (FIRST). Significant new surveys appeared at every wavelength, and continue to do so. We discuss a number of them in more detail in Sec. 4.

The early redshift observations by Vesto Melvin Slipher in the 1920's led to the discovery of the expanding universe by Edwin Hubble. Galaxy redshift surveys in a more modern sense can be dated from the pioneering work by Humason et al. (1956).

The early 1980's brought the first extensive redshift surveys, driven largely by the studies of the large-scale structure (LSS). An important milestone was the first Center for Astrophysics (CfA) redshift survey, conducted by Marc Davis, David Latham, John Huchra, John Tonry (Davis et al. 1982, Huchra et al. 1983), and their collaborators, based on the optical spectroscopy of ~ 2,300 galaxies down to $m_B \approx$ 14.5 mag, selected from the Zwicky and Nilson catalogs, obtained at Mt. Hopkins, Arizona. Arecibo redshift survey, conducted by Riccardo Giovanelli, Martha Haynes, and collaborators, used the eponymous radio telescope to measure ~ 2,700 galaxy redshifts through their H I 21-cm line. These surveys gave us the first significant glimpses of the LSS in the nearby universe. They were followed in a short order by the second CfA redshift survey, led by John Huchra and Margaret Geller (Geller & Huchra 1989), that covered galaxies down to $m_B \approx$ 15.5 mag; it was later combined with other survey for a total of ~ 18,000 redshifts, compiled in the *Updated Zwicky Catalog* (UZC; Falco et al. 1999). Additional H I surveys from Arecibo, added up to ~ 8,000 galaxies. Several redshift surveys used target selection on the basis of FIR sources detected by the IRAS satellite, in order to minimize the selection effects due to the interstellar extinction (Strauss et al. 1992, Fisher et al. 1995, Saunders et al. 2000). Many other redshift surveys of galaxies, obtained one redshift at a time, followed.

Together, these surveys added a few tens of thousands of galaxy redshifts, mainly used to study the LSS. Good reviews include, e.g., Giovanelli & Haynes (1991), Salzer & Haynes (1996), Lahav & Suto (2004), and Geller et al. (2011). As of this writing, John Huchra's ZCAT website is still maintained at https://www.cfa.harvard.edu/~dfabricant/huchra/zcat, and a good listing of redshift surveys is currently available at http://www.astro.ljmu.ac.uk/~ikb/research/galaxy-redshift-surveys.html.

Development of highly multiplexed (multi-fiber or multi-slit) spectrographs in the late 1980's and 1990's brought a new generation of massive redshift surveys, starting with the Las Campanas Redshift Survey, led by Steve Schectman, Gus Oemler, Bob Kirshner, and their collaborators (Shectman et al. 1996), that produced ~ 26,400 redshifts in sky strips covering ~ 700 deg$^2$. The field was changed dramatically in the late 1990's and 2000's by two massive redshift surveys, 2dF and SDSS, described in more detail below. Together, they contributed more than a million galaxy redshifts, and changed the field.

This blossoming of sky surveys created the exponential data explosion discussed in Sec. 1, and transformed the way astronomy is done, both scientifically and technologically. The state of the art at the transition from the photography era to the digital era is encapsulated well in the IAU Symposia 161 (MacGillivray et al. 1994) and 179 (McLean et al. 1998).

## 3. A Systematic Exploration of the Sky

Surveys are our main path towards a systematic exploration of the sky. They often follow an introduction of some new technology that allows us to observe the sky in some new way, e.g., in a previously unexplored wavelength regime, or to do so more efficiently, or more quickly. It is worth examining this process in a general sense. The discussion below follows in part the ideas first discussed by Harwit (1975); see also Harwit (2003). Harwit pointed out that fundamentally new discoveries are often made by opening of the new portions of the observable parameter space, and the key role of technology in enabling such advances. An interesting, complementary discussion was provided by Kurtz (1995).

Here we propose a framework for the representation of sky surveys (or, indeed, any astronomical observations), derived catalogs of measurements, and data exploration and analysis. The purpose is to have a way of quantitatively describing and comparing these large and complex data sets, and the process for their systematic exploration. We distinguish three types of data parameter spaces, for the observations, measured properties of detected sources in these observations, and physical (rather than apparent) properties of these sources.

In the context of exploration of large digital sky surveys and Virtual Observatory, some of these ideas have been proposed by Djorgovski et al. (2001abc, 2002). The present treatment develops further the concept of observational or data parameter space.

### *3.1 The Role of Technology*

Progress in astronomy has always been driven by technology. From the viewpoint of surveys, the milestone technologies include the development of astrophotography (late 1800's), Schmidt telescopes (1930's), radio electronics (1940's to the present), access to space (from 1960's onward), ubiquitous and inexpensive computing, and digital detectors in every wavelength regime, notably the CCDs (1980's to the present).

A question naturally arises, what is the next enabling technology? The same technology that gave us computers with capabilities increasing according to Moore's law, and digital detectors like the CCDs, viz. VLSI, also enabled the rise of the modern information and computation technology (ICT) in both hardware and software forms, the internet and the Web. Most of the computing now is not about number crunching and solving a lot of differential equations quickly (although of course we still do that), but it is about manipulation, processing, and searching of information, from the raw data to the high-level knowledge products. This is sometimes described as the Third (computation-intensive) and the Fourth (data-intensive) Paradigms of science, the First and the Second being the experiment and the analytical theory (Hey et al. 2009). Science progresses using all four of them.

The realm of key enabling technologies thus evolved historically from telescopes, to detectors, and now to information and computing technologies (ICT). In astronomy, the principal research environments in which this process unfolds incorporate the Virtual Observatory framework and, more broadly, Astroinformatics. These are reviewed elsewhere in these volumes. Equivalent situations exist in many other fields of science.

ICT is therefore the enabling technology behind the modern sky surveys and their scientific exploitation. Moore's law underlies both the exponential data growth, and the new methods needed to do science with these data. Surveys are already generating or enabling a lot of good science, and will dominate the data-rich astronomy at least in the first part of the 21$^{st}$ century.

*3.2 Data Parameter Spaces*

Every astronomical observation, surveys included, covers some finite portion of the *Observable Parameter Space* (OPS), whose axes correspond to the observable quantities, e.g., flux wavelength, sky coverage, etc. (see below). Every astronomical observation or a set thereof, surveys included, subtends a multi-dimensional volume (hypervolume) in this multi-dimentsional parameter space.

The OPS can be divided for convenience into four domains, roughly corresponding to the type of observations, although each of the four domains is really a limited projection of the OPS:

- The spectrophotometric domain, whose axes include the wavelength $\lambda$ (or the photon energy), flux $F_\lambda$ (that may be integrated over some finite bandpass), the spectroscopic resolution $R = \lambda/\Delta\lambda$, the inverse flux precision $F/\Delta F$ (so that the higher is better), and a polarization sub-domain that could consist, e.g., of the inverse precision of the measurements of Stokes parameters. One could divide it into spectroscopic and photometric sub-domains, but photometry can be seen as an extremely low resolution spectroscopy, and as the bandpasses shrink, the distinctions blur.
- The astrometric domain, whose axes include the pairs of coordinates (Equatorial, Ecliptic, Galactic…), and the astrometric accuracy $\Delta\theta$. The two coordinates can be collapsed into a net area coverage $\Omega$, if it doesn't matter where on the sky the observations are done; in this context, we use the word "area" to mean "solid angle", following the common, if somewhat misleading usage.
- The morphological domain, whose axes include the surface brightness $\mu$, and the angular resolution $\Delta\alpha$, that represents a "beam size", and should not be confused with the astrometric accuracy; given a sufficient S/N, it is possible to have $\Delta\theta \ll \Delta\alpha$.
- The time domain, whose axes include the time (say, the MJD or the UT) if it matters when the data are obtained, time baselines or sampling $\Delta t$, and the number of epochs (samples) $N_{exp}$ obtained at each $\Delta t$. These reflect the cadence and determine the window function of a survey.

Obviously, the four domains can share some of the axes and seldom make sense in isolation. An obvious example would be a single image: it covers a finite field of view, in some finite bandpass, with some limiting magnitude and a saturation level, and a limited angular resolution determined by the pixel size and/or the PSF. An imaging survey represents a sum of the OPS hyper-volumes of the individual images. Another example is the "cosmic haystack" of SETI searches, whose axes are the area, depth, and frequency (see, e.g., Tarter 2001).

We are also starting to add the non-electromagnetic information channels: neutrinos, gravity waves, and cosmic rays; they add their own dimensions to the general OPS.

The dimensionality of the OPS is given by the number of characteristics that can be defined for a given type of observations, although some of them may not be especially useful and could be ignored in a particular situation. For example, time domain axes make little sense for the observations taken in a single epoch. Along some axes, the coverage may be intrinsically discrete, rather than continuous. An observation can be just a single point along some axis, but have a finite extent in others. In some cases, polar coordinates may be more appropriate than the purely Cartesian ones.

Some parts of the OPS may be excluded naturally, e.g., due to quantum limits, diffraction limits, opacity and turbulence of the Earth's atmosphere or the Galactic ISM on some wavelengths, etc.; see Harwit (1975, 2003). Others are simply not accessible in practice, due to limitations of the available technology, observing time, and funding.

We can thus, in principle, measure a huge amount of information arriving from the universe, and so far we have sampled well only a relatively limited set of sub-volumes of the OPS in general, much better along some axes than others: we have fairly good coverage in the visible, NIR, and radio; more limited X-ray and FIR regimes; and very poor at higher energies still. The discrimination becomes sharper if we also consider their angular resolution, etc.

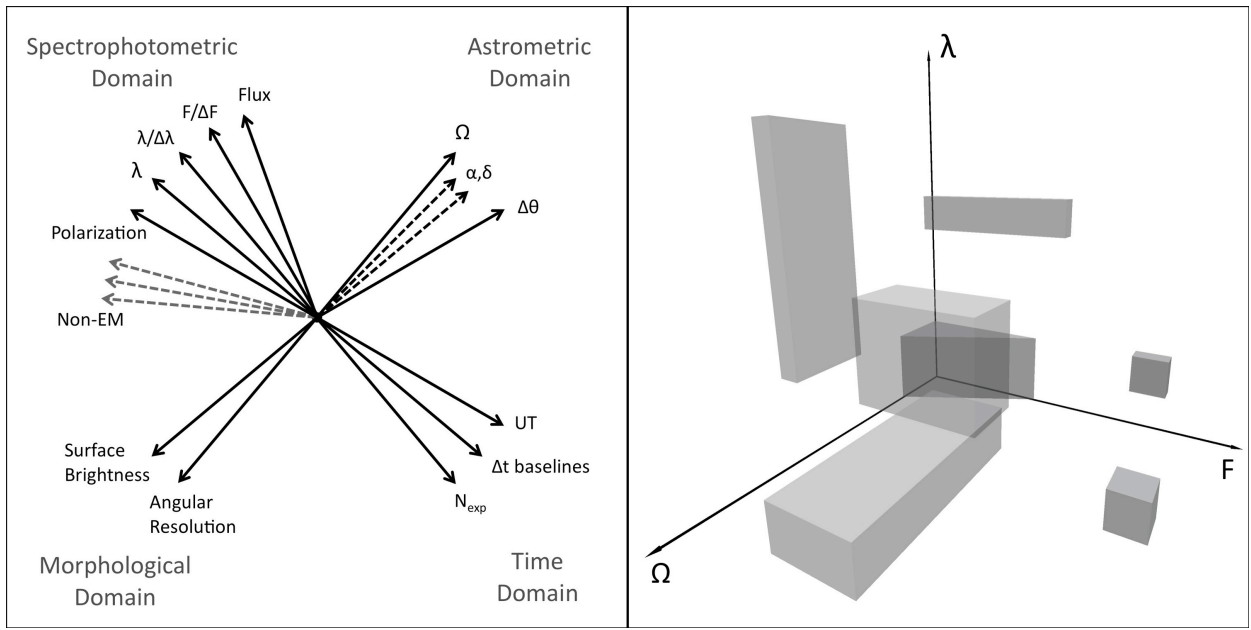

**Figure 1.** A schematic illustration of the Observable Parameter Space (OPS). All axes of the OPS corresponding to independent measurements are mutually orthogonal. Every astronomical observation, surveys included, carves out a finite hypervolume in this parameter space. *Left:* Principal axes of the OPS, grouped into four domains; representing such high-dimensionality parameter spaces on a 2D paper is difficult. *Right:* A schematic representation of a particular 3D representation of the OPS. Each survey covers some solid angle ($\Omega$), over some wavelength range ($\lambda$), and with some dynamical range of fluxes (F). Note that these regions need not have orthogonal, or even planar boundaries.

The coverage and the dimensionality of the OPS determine what *can* be detected or measured; it fully describes the scope and the limitations of our *observations*. It is in principle enormous, and as our observing technologies improve, we cover an ever greater portion of it. Selection effects due to the survey limitations are also more apparent in such a representation. Examining the coverage of the OPS can yield insights for the optimal strategies for future explorations.

Alternatively, knowing how given types of objects populate the OPS, one can optimize the coverage strategy to find them; for example, Supernovae, or brown dwarfs. We note that the inverse of the area coverage represents a limiting source density: if a survey covers the area $\Omega$, in order to detect $N$ sources of a given species, their surface density on the sky must be greater than $N/\Omega$. For example, down to $r < 20$ mag, the surface density of quasars at redshifts $z > 4$ is $\sim 4.1 \times 10^{-2}$ deg$^{-2}$; thus, in a survey with such limiting magnitude, in order to find $\sim 100$ such objects for subsequent studies, one has to cover at least $\sim 2,500$ deg$^2$.

Two relatively poorly explored domains of the OPS may be especially interesting: the low surface brightness universe, and the time domain, at any wavelength (Djorgovski et al. 2001ab, Brunner et al. 2001b, Diercks 2001). For example, the subject of possible missing large populations of low surface brightness galaxies has been debated extensively in the literature (see, e.g., Impey & Bothun 1997), but a systematic exploration of the low surface brightness universe is just starting. The time domain is currently undergoing a vigorous development, thanks to the advent of the modern synoptic sky surveys (Paczynski 2000, Djorgovski et al. 2012, and the volume edited by Griffin et al. 2012); we review some of them below.

As catalogs of sources and their measured properties are derived from imaging surveys, they can be represented as points (or vectors) in the *Measurement Parameter Space* (MPS). Every measured quantity for the individual sources has a corresponding axis in the MPS. Some could be derived from the primary measured quantities; for example, if the data are obtained in multiple bandpasses, we can form axes of flux ratios or colors; a difference of magnitudes in different apertures forms a concentration index; surface brightness profiles of individual objects can be constructed and parametrized, e.g., with the Sersic index; and so on. Some parameters may not even be representable as numbers, but rather as labels; for example, morphological types of galaxies, or a star vs. a galaxy classification.

While OPS represents the scope and the limitations of *observations,* MPS is populated by the detected *sources and their measured properties*. It describes completely the content of catalogs derived from the surveys.

Some of the axes of the OPS also pertain to the MPS; for example, fluxes or magnitudes. In those cases, one can map objects back to the appropriate projections of the OPS. However, although there is only one flux axis in a given photometric bandpass in the OPS, there may be several corresponding derived axes in the MPS, e.g., magnitudes measured in different apertures (obviously, such sets of measurement axes are correlated, and would not be mutually orthogonal). Some axes of the OPS would not have meaningful counterparts in the MPS; for example, the overall area coverage. There may also be axes of the MPS that are produced by measurements of images that are meaningless in the OPS, for example parameters describing the morphology of objects, like concentration indices, ellipticities, etc. For all of these reasons, it makes sense to separate the OPS from the MPS.

Each detected source is then fully represented as a feature vector in the MPS ("features" is a commonly used computer-science term for what we call measured parameters here). Modern imaging surveys may measure hundreds of parameters for each object, with a corresponding dimensionality of the MPS; however, many axes can be highly correlated (e.g., magnitudes in a series of apertures). It is thus a good idea to reduce the dimensionality of the MPS to a minimum set of independent axes, before proceeding to further data analysis.

The MPS in itself can be a useful research tool. For example, morphological classification of objects (stars vs. galaxies, galaxies of different types, etc.), as well as a detection and removal of measurement artifacts, which often appear as outliers from the bulk of the measurements, can be accomplished very effectively in the MPS. Sometimes, a judiciously chosen subset of parameters can be used for such tasks. Statistical and Machine Learning tools like the Kohonen Self-Organizing Maps (SOM; Kohonen 1989) can be used to find the most discriminating parameters for a given problem.

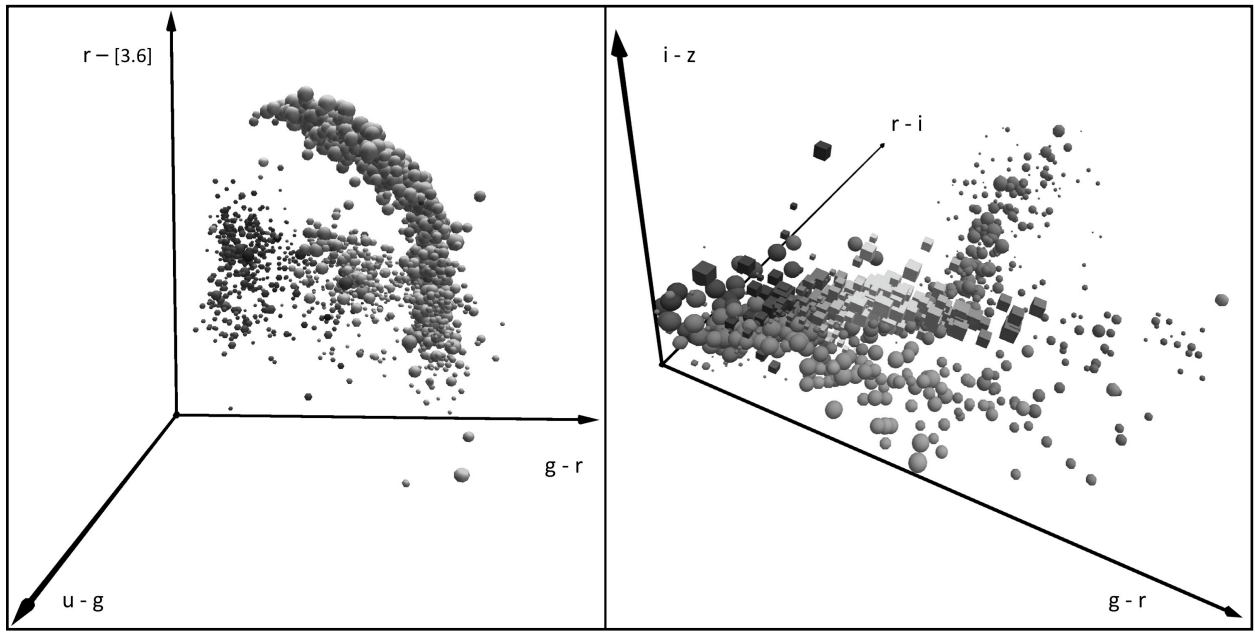

**Figure 2.** Examples of 3D projections of the Measurement Parameter Space (MPS), specifically colors. Only a couple of thousand of objects are plotted, for clarity. *Left:* Colors of quasars in the SDSS $u'g'r'$ and the *Spitzer* 3.6 μm band, from Richards et al. (2009). *Right:* Colors of stars, quasars and galaxies in the SDSS $u'g'r'i'z'$ color space. Objects of different types or sub-types form clusters and sequences in the MPS, that may be apparent in some of its projections, but not in others. In these examples, distributions in the space of observable parameters also carry a physical meaning, and can be used to separate the classes, or to search for rare types of objects, e.g., high-redshift quasars. Thus, visual inspection and exploration of data represented in the MPS can lead directly to new insights and discoveries. In addition to the 3 spatial axes, one can also use the sizes, shapes, and colors of the data points in order to encode additional dimensions. A more rigorous approach would be to use various data mining techniques, such as the clustering, classification, anomaly or outlier searches, correlation searches, and similar tasks.

Observed properties of detected sources can then be translated through the process of data reductions and analysis into their physical properties, usually requiring some additional or interpretative knowledge, e.g., distances, and/or assuming some data model. These form the *physical parameter space* (PPS), where typically the scientific discoveries are made. The MPS describes *observations* of individual sources found in a catalog; the PPS is populated by astronomical *objects*, again quantified as feature vectors. For example, an observed color-magnitude diagram for stars in some field is a 2-dimensional projection of the MPS; the corresponding temperature-luminosity diagram is a projection of the PPS. A magnitude belongs to the MPS; a luminosity is a feature of the PPS. There is of course no reason to stop at 2 dimensions, and in principle, the PPS can have at least as many axes as the MPS from which it is derived, although often many can be deliberately ignored in a given analysis. Also, the PPS can have additional axes of derived parameters, e.g., masses, chemical abundances (derived from the observed equivalent widths in the MPS), and so on.

The MPS and the PPS may partly overlap, typically with the axes that represent distance-independent quantities, for example colors or some other measures of the spectral energy distributions, such as the spectral indices, or hardness ratios; surface brightness; and measures of source morphology, e.g., concentration indices; or various measures of the variability, e.g., periods, if any, decay times, etc. (In a cosmological context, relativistic corrections have to be applied to the surface brightness and any time intervals.)

Surfaces delimiting the hyper-volumes covered by surveys in the OPS map into the selection effects in the MPS and its projections to lower dimensionality parameter spaces. These, in turn, map directly into the selection effects in the PPS. The combination of these parameter spaces thus represents a quantitative framework for data analysis and exploration.

*3.3 Exploring the Parameter Spaces*

An early vision of such a systematic exploration of the observable universe was promoted by Zwicky (1957), who was as usual far ahead of his time, dubbing it the "Morphological Box Approach". That was perhaps an unfortunate wording; reading Zwicky's writings, it is clear that what he meant is very much like the OPS and PPS defined above. Zwicky was convinced that a systematic, orderly exploration of these parameter spaces may lead to the discoveries of previously unknown phenomena. While Zwicky was limited by the observational technology available to him at the time, today we can bring these ideas to fruition, thanks to the thorough and systematic coverage of the sky in so many different ways by sky surveys.

Astronomical objects and phenomena are characterized by a variety of physical parameters, e.g., luminosities, spectral energy distributions, morphologies, or variability patterns. Sometimes, correlations among them exist, and their interpretation leads to a deeper physical understanding. Objects of different kinds, e.g., stars, quasars, galaxies of different kinds, etc., form clusters in the PPS, reflecting their physical distinctions, and that in itself represents some knowledge and understanding. Conversely, identifying such clusters (or outliers form them) in the space of physical properties can lead to discoveries of new types or sub-types of objects or phenomena.

It is important to note that the finite, specific hypervolumes of the OPS in any given survey are effectively "window functions" that determine what *can* be discovered: if a particular type of objects or phenomena avoids the probed region, then none will be found, and we don't know what we are missing in the portions of the OPS that were not covered. This argues for an OPS coverage, spread among a number of surveys, that covers as much of the accessible OPS hypervolume as possible and affordable.

The OPS and the PPS may overlap in some of the axes, and new discoveries may be made in the OPS alone, typically on the basis of the shapes of the spectral energy distributions. This generally happens when a new wavelength window opens up, e.g., some blue "stars" turned out to be a new phenomenon of nature, quasars, when they were detected as radio sources; and similar examples can be found in every other wavelength regime. The unexpectedness of such discoveries typically comes from some hidden assumptions, e.g., that all star-like objects will have a thermal emission in a certain range. When astronomical sources fail to meet our expectations, discoveries are made. Improvements in angular resolution or depth can also yield discoveries, simply by inspection. Of course, noticing something that appears new or unusual is just the first step, and follow-up observations and interpretative analysis are needed.

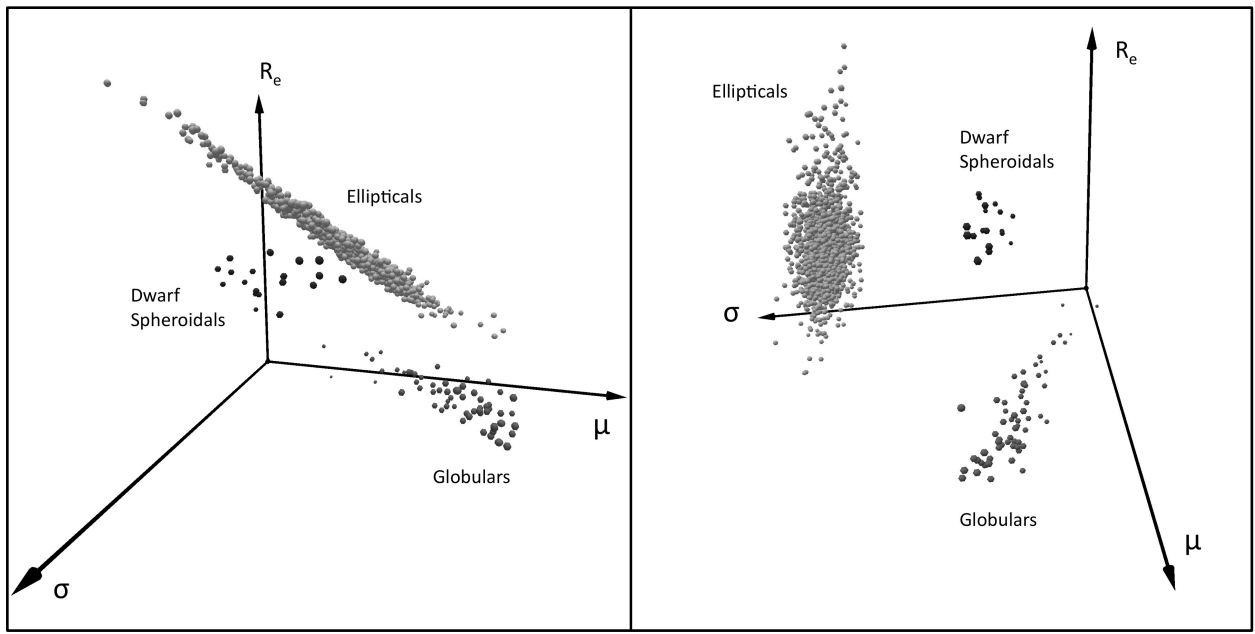

**Figure 3.** A simple illustration of a physical parameter space (PPS), viewed from two different angles. Families of dynamically hot stellar systems are shown in the parameter space whose axes are logs of effective radii, mean surface brightness, and the central velocity dispersion in he case of ellipticals, and the maximum rotational speed in the case of spirals. In the panel on the left, we see an edge-on view of the Fundamental Plane (FP) for elliptical galaxies (Djorgovski & Davis 1987). On the right we see a clear separation of different families of objects, as first noted by Kormendy (1985). They form distinct clusters, some of which represent correlations of their physical parameters that define them as families in an objective way. This galaxy parameter space (Djorgovski 1992ab) is a powerful framework for the exploration of such correlations, and their implications for galaxy evolution. Data for ellipticals are from La Barbera et al. (2009), and for dwarf spheroidals and globulars are compiled from the literature.

Sometimes we expect to find objects in some region of the parameter space. For example, most quasars (especially the high-redshift ones) and brown dwarfs are found in particular regions of the color space, on the basis of an assumed (correct) model of their spectral energy distributions.

Thus, PPS approach can thus be put to a good use for a *directed discovery*, to form samples of known, interesting types of objects by introduction of a prior knowledge. However, it also enables discoveries that were unanticipated, and approach we may call an *organized serendipity*.

The PPS is *not* uniformly populated: astronomical objects cluster and correlate in particular regions, for good physical reasons, and are likewise absent from most of the PPS hypervolume. For example, only a few narrow, well-defined sequences are populated in the HR diagram, reflecting the physics of stellar structure and evolution. Some of these clusters correspond to the known classes of objects, but in principle other, as yet unknown clusters may exist in the unexplored corners of the PPS. Both the existence of clusters in the OPS, and the empty regions thereof in principle contain valuable information about the cosmic phenomena. Therein lies the knowledge discovery potential of this approach.

Various data mining (DM) techniques, such as supervised or unsupervised clustering analysis, outlier and anomaly searches, multivariate analysis, can be used to map out the PPS and thus the kinds of the astronomical objects that populate it (e.g., de Carvalho et al. 1995, Yoo et al. 1996). If applied in the catalog domain, the data can be viewed as a set of $N$ points or vectors in a $D$-dimensional parameter space, where $N$ can be in the range of many millions or even billions, and $D$ in the range of a few tens to many hundreds. The data may be clustered in $k$ statistically distinct classes, that may be uncovered through the DM process. This is computationally and algorithmically a highly non-trivial problem, that is now being addressed in the Astroinformatics and related disciplines. An example of a data exploration system along these lines is *Data Mining and Exploration* (DAME; Brescia et al. 2010, 2012; http://dame.dsf.unina.it). A caveat is in order: most of the commonly used DM techniques implicitly or explicitly assume that the data are complete, error-free, and free of artifacts, none of which is true in reality. Thus, they cannot be applied blindly, and scientists should understand their limitations when using them.

Effective visualization techniques can go a long way towards an exploration and intuitive understanding of all of these parameter spaces. Representation of highly-dimensional parameter spaces in 3D projections (which in themselves tend to be projected to 2D surfaces, such as the paper or a screen) poses many challenges of its own, as we are biologically limited to see in 3 dimensions. Various tricks can be used to encode up to a dozen dimensions in a pseudo-3D plot, but there is no reason why the nature would limit itself to interesting or meaningful structures in a small number of dimensions.

However, not all parameters may be equally interesting or discriminating, and lowering the dimensionality of the PPS to some more appropriate, judiciously chosen subset of parameters may be necessary. This is where the scientists' skill comes into the play.

Discovery of particularly rare types of objects, whether known or unknown, can be done as a search for outliers in a judiciously chosen projection of the PPS. Such things might have been missed so far either because they are rare, or because they would require a novel combination of data or a new way of looking at the data. A thorough, large-scale, unbiased, multi-wavelength census of the universe will uncover them, if they do exist (and surely we have not yet found all there is out there). Surveys that detect large numbers of sources are the only practical way to find extremely rare ones; for example, down to the ~20 mag, approximately one in a million star-like objects at high Galactic latitude is actually a quasar at $z > 4$, and that is not even a particularly rare type of an object these days.

Sometimes new phenomena are found (or rediscovered) independently in different regions of the OPS. Whether such repeated findings imply that there is a finite (and even a relatively modest) number of fundamentally different astronomical phenomena in the universe is an interesting question (Harwit & Hildebrand 1986).

Most of the approaches described so far involve searches in some parameter or feature space, i.e., catalogs derived from survey images. However, we can also contemplate a direct exploration of sky surveys in the image (pixel) domain. Automated pattern recognition and classification tools can be used to discover sources with a particular image morphology (e.g., galaxies of a certain type). An even more interesting approach would be to employ machine learning and artificial intelligence techniques to search through panoramic images (perhaps matched from multiple wavelengths) for unusual image patterns; known example may be the gravitationally lensed arcs around rich clusters, or supernova remnants.

Finally, an unsupervised classification search for unusual patterns or signals in astronomical data represents a natural generalization of the SETI problem (Djorgovski 2000). Manifestations of ETI may be found as "anomalous" signals and patterns in some portion of the OPS that has not been well explored previously.

## 4. Characteristics and Examples of Sky Surveys

### 4.1 A Sampler of Panoramic Sky Surveys and Catalogs

Sky surveys, large and small, are already too numerous to cover extensively here, and more are produced steadily. In the Appendix, we provide a listing, undoubtedly incomplete, of a number of them, with the Web links that may be a good starting point for a further exploration. Here we describe some of the more popular panoramic (wide-field, single-pass) imaging surveys, largely for the illustrative purposes, sorted in wavelength regimes, and in a roughly chronological order.

However, our primary aim is to illustrate the remarkable richness and diversity of the field, and a dramatic growth in the data volumes, quality, and homogeneity. Surveys are indeed the backbone of contemporary astronomy.

Table 1 lists basic parameters for a number of popular surveys, at least as of this writing. It is meant as illustrative, rather than complete in any way. In the Appendix, we list a much larger number of the existing surveys and their websites, from which the equivalent information can be obtained. We also refrain from talking too much about the surveys that are still in the planning stages; the future will tell how good they will be.

### 4.1.1 Surveys and Catalogs in the Visible Regime

The POSS-II photographic survey plate scans obtained at the Space Telescope Science Institute (STScI), led to the *Digitized Palomar Observatory Sky Survey* (DPOSS; Djorgovski et al. 1994, 1997a, 1999, Weir et al. 1995a; http://www.astro.caltech.edu/~george/dposs), a digital survey of the entire Northern sky in three photographic bands, calibrated by CCD observations to SDSS-like *gri*. The plates are scanned with 1.0 arcsec pixels, in rasters of 23,040 square, with 16 bits per pixel, producing about 1 GB per plate, or about 3 TB of pixel data in total. They were processed independently at STScI for the purposes of constructing a new guide star catalog for the HST (Lasker et al. 2008), and at Caltech for the DPOSS project. Catalogs of all the detected objects on each plate were generated, down to the flux limit of the plates, which roughly corresponds to the equivalent limiting magnitude of $m_B \sim 22$ mag. A specially developed software package, *Sky Image Cataloging and Analysis Tool* (SKICAT; Weir et al. 1995c) was used to process and analyze the data. Machine learning tools (decision trees and neural nets) were used for star-galaxy classification (Weir et al. 1995b, Odewahn et al. 2004), accurate to 90% down to ~ 1 mag above the detection limit. An extensive program of CCD calibrations was done at the Palomar 60-inch telescope (Gal et al. 2005). The resulting object catalogs were combined and stored in a relational database. The DPOSS catalogs contain ~ 50 million galaxies, and ~ 500 million stars, with over 100 attributes measured for each object. They generated a number of studies, e.g., a modern, objectively generated analog of the old Abell cluster catalog (Gal et al. 2009).

| Survey | Type | Duration | Bandpasses | Lim. flux | Area coverage | $N_{sources}$ | Notes |
|---|---|---|---|---|---|---|---|
| DSS scans | Visible | 1950's-1990's | $B$ (~ 450 nm)<br>$R$ (~ 650 nm)<br>$I$ (~ 800 nm) | 21 – 22 mag<br>20 – 21 mag<br>19 – 20 mag | Full sky | ~ $10^9$ | Scans of plates from the POSS and ESO/SERC surveys |
| SDSS-I<br>SDSS-II<br>SDSS-III | Visible | 2000-2005<br>2005-2008<br>2009-2014 | $u$ (~ 800 nm)<br>$g$ (~ 800 nm)<br>$r$ (~ 800 nm)<br>$i$ (~ 800 nm)<br>$z$ (~ 800 nm) | 22.0 mag<br>22.2 mag<br>22.2 mag<br>21.3 mag<br>20.5 mag | 14,500 deg$^2$ | 4.7 × $10^8$ | Numbers quoted for DR8 (2011). In addition, spectra of ~ 1.6 million objects |
| 2MASS | Near IR | 1997-2001 | $J$ (~ 1.25 μm)<br>$H$ (~ 1.65 μm)<br>$K_s$ (~ 2.15 μm) | 15.8 mag<br>15.1 mag<br>14.3 mag | Full sky | 4.7 × $10^8$ | |
| UKIDSS | Near IR | 2005-2012 | $Y$ (~ 1.05 μm)<br>$J$ (~ 1.25 μm)<br>$H$ (~ 1.65 μm)<br>$K$ (~ 2.2 μm) | 20.5 mag<br>20.0 mag<br>18.8 mag<br>18.4 mag | 7,500 deg$^2$ | ~ $10^9$ | Estim. final numbers quoted for the LAS; deeper surveys over smaller areas also done |
| IRAS | Mid/Far IR (space) | 1983-1986 | 12 μm<br>25 μm<br>60 μm<br>100 μm | 0.5 Jy<br>0.5 Jy<br>0.5 Jy<br>1.5 Jy | Full sky | 1.7 × $10^5$ | |
| NVSS | Radio | 1993-1996 | 1.4 GHz | 2.5 mJy | 32,800 deg$^2$ | 1.8 × $10^6$ | Beam ~ 45 arcsec |
| FIRST | Radio | 1993-2004 | 1.4 GHz | 1 mJy | 10,000 deg$^2$ | 8.2 × $10^5$ | Beam ~ 5 arcsec |
| PMN | Radio | 1990 | 4.85 GHz | ~ 30 mJy | 16,400 deg$^2$ | 1.3 × $10^4$ | Combines several surveys |
| GALEX | UV (space) | 2003-2012 | 135 – 175 nm<br>175 – 275 nm | 20.5 mag AIS<br>23 mag MIS | AIS 29,000<br>MIS 3,900 | 6.5 × $10^7$<br>1.3 × $10^7$ | As of GR6 (2011); also some deeper data |
| Rosat | X-ray (space) | 1990-1999 | 0.04 – 2 keV | ~ $10^{-14}$ erg cm$^{-2}$ s$^{-1}$ | Full sky | 1.5 × $10^5$ | Deeper, small area surveys also done |
| Fermi LAT | γ-ray (space) | 2008-? | 20 MeV to 30 GeV | 4 × $10^{-9}$ erg cm$^{-2}$ s$^{-1}$ | Full sky | ~ 2 × $10^3$ | LAT instrument only; in addition, GRBM |

**Table 1.** Basic properties of some of the popular wide-field surveys. The quoted limiting magnitudes are on the Vega zero-point system for DSS, 2MASS, and UKIDSS; on the AB$_\nu$ system for GALEX; and SDSS zero-points are close to the AB$_\nu$ system. See the text for more details and websites.

The *United States Naval Observatory Astrometric Catalog* (USNO-A2, and an improved version, USNO-B1, http://www.nofs.navy.mil/data/FchPix) were derived from the POSS-I, POSS-II and ESO/SERC Southern sky survey plate scans, done using the Precision Measuring Machine (PMM) built and operated by the United States Naval Observatory Flagstaff Station (Monet et al. 2003). They contain about a billion sources over the entire sky down to a limiting magnitude of $m_B \sim 20$ mag, and they are now commonly used to derive astrometric solutions for optical and NIR images. A combined astrometric data set from several major modern catalogs is currently being served by the USNO's *Naval Observatory Merged Astrometric Dataset* (NOMAD; http://www.usno.navy.mil/USNO/astrometry/optical-IR-prod/nomad).

The *Sloan Digital Sky Survey* (SDSS; Gunn et al. 1998, 2006, York et al. 2000, Fukugita et al. 1996; http://sdss.org) is a large international astronomical collaboration focused on constructing the first CCD photometric survey high Galactic latitudes, mostly in the Northern sky. The survey and its extensions (SDSS-II and SDSS-III) eventually covered $\sim 14{,}500$ deg$^2$, i.e., more than a third of the entire sky. A dedicated 2.5 m telescope at Apache Point, New Mexico, was specially designed to take wide field (3°×3°) images using a mosaic of thirty 2048×2048 pixel CCDs, with 0.396 arcsec pixels, and with additional devices used for astrometric calibration. The imaging data were taken in the drift scanning mode along great circles on the sky. The imaging survey uses 5 passbands ($u'g'r'i'z'$), with limiting magnitudes of 22.0, 22.2, 22.2, 21.3, and 20.5 mag, in the $u'g'r'i'z'$ bands respectively. The spectroscopic component of the survey is described below. The total raw data volume is several tens of TB, with an at least comparable amount of derived data products.

The initial survey (SDSS-I, 2000-2005) covered $\sim 8{,}000$ deg$^2$. SDSS-II (2005-2008) consisted of *Sloan Legacy Survey*, that expanded the area coverage to $\sim 8{,}400$ deg$^2$, and catalogued 230 million objects, the *Sloan Extension for Galactic Understanding and Exploration* (SEGUE) that obtained almost a quarter million spectra over $\sim 3{,}500$ deg$^2$, largely for the purposes of studies of the Galactic structure, and the *Sloan Supernova Survey* which confirmed 500 Type Ia SNe. The latest survey extension, SDSS-III (2008-2014), is still ongoing at this time.

The data have been made available to the public through a series of well-documented releases, the last one (as of this writing), DR8, occurred in 2011. In all, by 2011, photometric measurements for $\sim 470$ million unique objects have been obtained, and $\sim 1.6$ million spectra.

More than perhaps any other survey, SDSS deserves the credit for transforming the culture of astronomy in regards to sky surveys: they are now regarded as fundamental data sources with a rich information content. SDSS enabled – and continues to do so – a vast range of science, not just by the members of the SDSS team (of which there are many), but also the rest of the astronomical community, spanning topics from asteroids to cosmology. Due to its public data releases, SDSS may be the most productive astronomical project so far.

The *Panoramic Survey Telescope & Rapid Response System* (Pan-STARRS; http://pan-starrs.ifa.hawaii.edu; Kaiser et al. 2002) is a wide-field panoramic survey developed at the University of Hawaii's Institute for Astronomy, and operated by an international consortium of institutions. It is envisioned as a set of four 1.8 m telescopes with a 3° field of view each, observing the same region of the sky simultaneously, each equipped with a $\sim 1.4$ Gigapixel CCD camera with 0.3 arcsec pixels. As of this writing, only one of the telescopes (PS1) is operational, at Haleakala Observatory on Maui, Hawaii; it started science operations in 2010. The survey may be transferred to the Mauna Kea Observatory, if additional telescopes are built. The CCD cameras, consisting of an array of 64, 600×600 pixel CCDs, incorporate a novel technology, Orthogonal Transfer CCDs, that can be used to compensate the atmospheric tip-tilt distortions (Tonry et al. 1997), although that feature has not yet been used for the regular data taking. PS1 can cover up to 6,000 deg$^2$ per night and generate up to several TB of data per night, but not all images are saved, and the expected final output is estimated to be ~ 1 PB per year.

The primary goal of the PS1 is to survey ~ ¾ of the entire sky (the "3π" survey) with 12 epochs in each of the 5 bands (*grizy*). The coadded images should reach considerably deeper than SDSS. A dozen key projects, some requiring additional observations, are also under way. The survey also has a significant time-domain component. The data access is restricted to the consortium members until the completion of the "3π" survey.

Its forthcoming Southern sky counterpart is *SkyMapper* (http://msowww.anu.edu.au/skymapper), developed by Brian Schmidt and collaborators at the Australian National University. It is undergoing the final commissioning steps as of this writing, and it will undoubtedly be a significant contributor when it starts full operations. The fully automated 1.35 m wide-angle telescope is located at Siding Spring Observatory, Australia. It has a ~ 270 Megapixel CCD mosaic camera that covers ~ 5.7 deg$^2$ in a single exposure, with 0.5 arcsec pixels. *SkyMapper* will cover the entire southern sky 36 times in 6 filters (SDSS *g′r′i′z′*, a Strömgren system-like *u*, and a unique narrow *v* band near 4000Å). It will generate ~100 MB of data per second during every clear night, totaling about 500 Terabytes of data at the end of the survey. A distilled version of the survey data will be made publicly available. Its scientific goals include: discovering dwarf planets in the outer solar system, tracking asteroids, create a comprehensive catalog of the stars in our Galaxy, dark matter studies, and so on.

Another Southern sky facility is the *Very Large Telescope (VLT) Survey Telescope* (VST; http://www.eso.org/public/teles-instr/surveytelescopes/vst.html) that is just starting operations as of this writing (2011). This 2.6 m telescope covers a ~1 deg$^2$ field, with a ~ 270 Megapixel CCD mosaic consisting of 32, 2048×4096 pixel CCDs. The VST ATLAS survey will cover 4,500 deg$^2$ of the Southern sky in five filters (*ugriz*) to depths comparable to those of the SDSS. A deeper survey, KIDS, will reach ~ 2 – 3 mag deeper over 1,500 deg$^2$ in *ugri* bands. An additional survey, VPHAS+, will cover 1,800 deg$^2$ along the Galactic plane in *ugri* and Hα bands, to depths comparable to the ATLAS survey. These and other VST surveys will also be complemented by near-infrared data from the VISTA survey, described below. They will provide an imaging base for spectroscopic surveys by the VLT. Scientific goals include studies of dark matter and dark energy, gravitational lensing, quasar searches, structure of our Galaxy, etc.

An ambitious new *Dark Energy Survey* (DES; http://www.darkenergysurvey.org) should start in 2012, using the purpose-built the *Dark Energy Camera* (DECam), on the Blanco 4 m telescope at CTIO. It is a large international collaboration, involving 23 institutions. DECam is a 570 Megapixel camera with 74 CCDs, with a 2.2 deg$^2$ FOV. The primary goal is to catalog 5,000 deg$^2$ in the Southern hemisphere, in *grizy* bands, in an area overlaps with several other surveys, for a maximum leveraging. It is expected to see over 300 million galaxies during the 525 nights, and find and measure ~ 3,000 SNe, and use several observational probes of dark energy in order to constrain its physical nature.

The GAIA mission (http://gaia.esa.int) is expected to be launched in 2012, expected to revolutionize astronomy and provide fundamental data for a better understanding of the structure of our Galaxy, distance scale, stellar evolution, and other topics. It will measure positions with a 20 microarcsec accuracy for about a billion stars (i.e., at the depth roughly matching the POSS surveys), from which parallaxes and proper motions will be derived. In addition, it will obtain low resolution spectra for many objects. It is expected to cover the sky ~ 70 times over a projected 5 year mission, and thus discover numerous variable and transient sources.

Many other modern surveys in the visible regime invariably have a strong synoptic (time domain) component, reflecting some of their scientific goals. We describe some of them below.

### 4.1.2 Surveys in the Infrared

The *Two Micron All-Sky Survey* (2MASS; 1997-2002; http://www.ipac.caltech.edu/2mass; Skrutskie et al. 2006) is a near-IR (*J, H*, and $K_S$ bands) all-sky survey, done as a collaboration between the University of Massachusetts, which constructed the observatory facilities and operated the survey, and the Infrared Processing and Analysis Center (IPAC) at Caltech, which is responsible for all data processing and archives. It utilized two highly automated 1.3 m telescopes, one at Mt. Hopkins, AZ and one at CTIO, Chile. Each telescope was equipped with a three-channel camera with HgCdTe detectors, and was capable of observing the sky simultaneously at *J, H*, and $K_S$ with 2 arcsec pixels. It remains as the only modern, ground-based sky survey that covered the entire sky. The survey generated ~ 12 TB of imaging data, a point source catalog of ~ 300 million objects, with a 10-σ limit of $K_S$ < 14.3 mag (Vega zero point), and a catalog of ~ 500,000 extended sources. All of the data are publicly available via a web interface to a powerful database system at IPAC.

The *United Kingdom Infrared Telescope (UKIRT) Infrared Deep Sky Survey* (UKIDSS, 2005-2012, Lawrence et al. 2007, http://www.ukidss.org) is a NIR sky survey. Its wide-field component covers about 7,500 deg$^2$ of the Northern sky, extending over both high and low Galactic latitudes, in the *YJHK* bands down to the limiting magnitude of *K* = 18.3 mag (Vega zero point). UKIDSS is made up of five surveys and includes two deep extra-Galactic elements, one covering 35 deg$^2$ down to K = 21 mag, and the other reaching K = 23 mag over 0.77 deg$^2$. The survey instrument is WFCAM on the UK Infrared Telescope (UKIRT) in Hawaii. It has four 2048x2048 IR arrays; the pixel scale of 0.4 arcsec gives an exposed solid angle of 0.21 deg$^2$. Four of the principal scientific goals of UKIDSS are: finding the coolest and the nearest brown dwarfs, studies high-redshift dusty starburst galaxies, elliptical galaxies and galaxy clusters at redshifts 1 < *z* < 2, and the highest-redshift quasars, at z ~ 7. The survey succeeded on all counts. The data were made available to the entire ESO community immediately they are entered into the archive, followed by a world-wide release 18 months later.

The *Visible and Infrared Survey Telescope for Astronomy* (VISTA; http://www.vista.ac.uk) is a 4.1 m telescope at the ESO's Paranal Observatory. Its 67 Megapixel IR camera uses 16 IR detectors, covers a FOV ~ 1.65° with 0.34 arcsec pixels, in *ZYJHK$_s$* bands, and a 1.18 μm narrow band. It will conduct a number of distinct surveys, the widest being the *VISTA Hemisphere Survey* (VHS), covering the entire Southern sky down to the limiting magnitudes ~ 20 – 21 mag in *YJHK$_s$* bands, and the deepest being the Ultra-VISTA, that will cover a 0.73 deg$^2$ field down to the limiting magnitudes ~ 26 mag in the *YJHK$_s$* bands, and the 1.18 μm narrow band. VISTA is expected to produce data rates exceeding 100 TB/year, with the publicly available data distributed through the ESO archives.

A modern, space-based all-sky survey is *Wide-field Infrared Survey Explorer* (WISE; 2009-2011; Wright et al. 2010; http://wise.ssl.berkeley.edu). It mapped the sky at 3.4, 4.6, 12, and 22 μm bands, with an angular resolution of 6.1, 6.4, 6.5, and 12.0 arcsec in the four bands, the number of passes depending on the coordinates. It achieved 5-σ point source sensitivities of about 0.08, 0.11, 1 and 6 mJy, respectively, in unconfused regions. It detected over 250 million objects, with over 2 billion individual measurements, ranging from near-Earth asteroids, to brown dwarfs, ultraluminous starbursts, and other interesting types of objects. Given its uniform, all-sky coverage, this survey is likely to play an important role in the future. The data are now publicly available.

Two examples of modern, space-based IR surveys over the limited areas include the *Spitzer Wide-area InfraRed Extragalactic* survey (SWIRE; http://swire.ipac.caltech.edu; Lonsdale et al. 2003), that covered 50 deg$^2$ in 7 mid/far-IR bandpasses, ranging from 3.6 μm to 160 μm, and the *Spitzer Galactic Legacy Infrared Mid-Plane Survey Extraordinaire* (GLIMPSE; http://www.astro.wisc.edu/sirtf/), that covered ~ 240 deg$^2$ of the Galactic plane in 4 mid-IR bandpasses, ranging from 3.6 μm to 8.0 μm. Both were followed up at a range of other wavelengths, and provided new insights into obscured star formation in the universe.

### 4.1.3 Surveys and Catalogs in the Radio

The *Parkes Radio Sources* (PKS, http://archive.eso.org/starcat/astrocat/pks.html) is a radio catalog compiled from major radio surveys taken with the Parkes 64 m radio telescope at 408 MHz and 2700 MHz over a period of almost 20 years. It consists of 8,264 objects covering all the sky south of Dec = +27° except for the Galactic Plane and Magellanic Clouds. Discrete sources were selected with flux densities in excess of ~50 mJy. It has been since been supplemented (PKSCAT90) with improved positions, optical identifications and redshifts as well as flux densities at other frequencies to give a frequency coverage of 80 MHz – 22 GHz.

Numerous other continnum surveys have been performed. An example is *Parkes-MIT-NRAO* (PMN, http://www.parkes.atnf.csiro.au/observing/databases/pmn/pmn.html) radio continuum surveys were made with the NRAO 4.85 GHz seven-beam receiver on the Parkes telescope during June and November 1990. Maps covering four declination bands were constructed from the survey scans with the total survey coverage over –87.5° < Dec < +10°. A catalog of 50,814 discrete sources with angular sizes approximately less than 15 arcmin, and a flux density greater than about 25 mJy (the flux limit varies for each band, but is typically this value) was derived from these maps. The positional accuracy is close to 10 arcsec in each coordinate. The 4.85 GHz weighted source counts between 30 mJy and 10 Jy agree well with previous results.

The *National Radio Astronomical Observatory* (NRAO), *Very Large Array* (VLA) *Sky Survey* (NVSS; Condon et al. 1998; http://www.cv.nrao.edu/nvss) is a publicly available, radio continuum survey at 1.4 GHz, covering the sky north of Dec = –40° declination, i.e., ~ 32,800 deg$^2$. The principal NVSS data products are a set of 2326 continuum map data cubes, each covering an area of 4°×4° with three planes containing the Stokes I, Q, and U images, and a catalog of discrete sources in them. Every large image was constructed from more than one hundred of the smaller, original snapshot images (over 200,000 of them in total). The survey catalog contains over 1.8 millions discrete sources with total intensity and linear polarization image measurements (Stokes I, Q, and U), with a resolution of 45 arcsec, and a completeness limit of about 2.5 mJy. The NVSS survey was performed as a community service, with the principal data products being released to the public by anonymous FTP as soon as they were produced and verified.

The *Faint Images of the Radio Sky at Twenty-cm* (FIRST; http://sundog.stsci.edu; Becker et al. 1995) is publicly available, radio snapshot survey performed at the NRAO VLA facility in a configuration that provides higher spatial resolution than the NVSS, at the expense of a smaller survey footprint. It covers ~ 10,000 deg$^2$ of the North and South Galactic Caps with a ~ 5 arcsec resolution. A final atlas of radio image maps with 1.8 arcsec pixels is produced by coadding the twelve images adjacent to each pointing center. The survey catalog contains around one million sources with a resolution of better than 1 arcsec. A source catalog including peak and integrated flux densities and sizes derived from fitting a two-dimensional Gaussian to each source is generated from the atlas. Approximately 15% of the sources have optical counterparts at the limit of the POSS-I plates; unambiguous optical identifications (< 5% false detection rates) are achievable to V ~ 24 mag. The survey area has been chosen to coincide with that of the SDSS. At the magnitude limit of the SDSS, approximately 50% of the FIRST sources have optical counterparts.

Two examples of modern H I 21 cm line surveys are the *HI Parkes All Sky Survey* (HIPASS; http://aus-vo.org/projects/hipass, http://www.atnf.csiro.au/research/multibeam/release/) and the Arecibo Legacy Fast ALFA survey (ALFALFA; http://egg.astro.cornell.edu; Giovanelli et al. 2005). These surveys cover the redshifts up to $z \sim 0.05 - 0.06$, detecing tens of thousands H I selected galaxies, high-velocity clouds, etc.

The new generation of ground-based radio sky surveys is intrinsically synoptic in nature; we discuss some of them below. A number of experiments are also addressing the epoch of reionization. The currently ongoing *Planck* mission (http://www.rssd.esa.int/Planck), in addition to its primary scientific goals of CMB-based cosmology, will also represent an excellent radio survey at high frequencies.

### 4.1.4 Surveys at Higher Energies

Numerous space missions have surveyed the sky at wavelengths ranging from UV to γ-rays. Some of the more modern, panoramic ones include:

*Galaxy Evolution Explorer* (GALEX; Martin et al. 2005; http://www.galex.caltech.edu), launched in 2003, is the first (and so far the only) nearly-all-sky survey UV mission. Data are obtained in two bands (1350–1750 and 1750–2750 Å) using microchannel plate detectors that provide time-resolved imaging. The All-sky Imaging Survey (AIS; it excludes some fields containing bright stars) reaches the depth comparable to the POSS surveys ($m_{AB} \simeq 20.5$ mag); a medium depth survey covers ~3,900 deg$^2$ down to $m_{AB} \simeq 23$ mag, and a deep imaging survey covers ~350 deg$^2$ down to $m_{AB} \simeq 25$ mag, In addition, a survey of several hundred nearby galaxies and a low-resolution (R = 100–200) slitless grism spectroscopic survey was performed over a limited area. The data are released in roughly yearly intervals, with the final one (GR7) expected in 2012.

ROSAT (Röntgensatellit; 1990-1999; http://www.dlr.de/dlr/en/desktopdefault.aspx/tabid-10424/) was a German Aerospace Center-led satellite X-ray telescope, with instruments built by Germany, UK and US. It carried an imaging X-ray Telescope (XRT) with three focal plane instruments: two Position Sensitive Proportional Counters (PSPC), a High Resolution Imager (HRI), and a wide field camera (WFC) extreme ultraviolet telescope co-aligned with the XRT that covered the wave band between and 6 angstroms (0.042 to 0.21 keV). The X-ray mirror assembly was a grazing incidence four-fold nested Wolter I type telescope with an 84 cm diameter aperture and 240 cm focal length. The angular resolution was less than 5 arcsec at half energy width. The XRT assembly was sensitive to X-rays between 0.1 to 2 keV.

The *ROSAT All Sky Survey* (Voges et al. 1999) was publicly released in 2000, and it is currently available from the MPE archive and the HEASARC archive at GSFC. It consists of 1378 distinct fields, with the bulk of all sky survey obtained by PSPC-C in scanning mode. The telescope detected more than 150,000 X-ray sources, ~ 20 times more than were previously known. The survey enabled a detailed morphology of supernova remnants and clusters of galaxies; detection of shadowing of diffuse X-ray emission by molecular clouds; detection of isolated neutron stars; discovery of X-ray emission from comets; and many other results.

## *4.2 Synoptic Sky Surveys and Exploration of the Time Domain*

A systematic exploration of the time domain in astronomy is now one of the most exciting and rapidly growing areas of astrophysics, and a vibrant new observational frontier (Paczynski 2000). A number of important astrophysical phenomena can be discovered and studied *only* in the time domain, ranging from exploration of the Solar System to cosmology and extreme relativistic phenomena. In addition to the studies of known, interesting time domain phenomena, e.g., supernovae and other types of cosmic explosions, there is a real and exciting possibility of discovery of new types of objects and phenomena. As already noted, opening new domains of the observable parameter space often leads to new and unexpected discoveries.

The field has been fueled by the advent of the new generation of digital synoptic sky surveys, which cover the sky many times, as well as the ability to respond rapidly to transient events using robotic telescopes. This new growth area of astrophysics has been enabled by information technology, continuing evolution from large panoramic digital sky surveys, to panoramic digital cinematography of the sky, opening the great new discovery space (Djorgovski et al. 2012).

Numerous surveys, studies, and experiments have been conducted in this arena, are in progress now, or are in planning stages, indicating the growing interest in time-domain astronomy, leading to the Large Synoptic Survey Telescope (LSST; Tyson et al. 2002, Ivezic et al. 2008). Today's surveys are opening the time domain frontier, yielding exciting results, and can be used to refine the scientific and methodological strategies for the future.

The key to progress in this emerging arena of astrophysics is the availability of substantial event data streams generated by panoramic digital synoptic sky surveys, coupled with a rapid follow-up of potentially interesting events (photometric, spectroscopic, and multi-wavelength). This, in turn, requires a rapid, real-time processing of massive sky survey data streams, with event detection, filtering, characterization, and rapid dissemination to the astronomical community. Over the past few years, we have been developing both synoptic sky survey data streams, and cyber-infrastructure needed for the rapid distribution and characterization of transient events and phenomena.

As in the previous subsection, here we describe some of the recent and ongoing synoptic sky surveys. This list is meant as illustrative, rather than complete, and the interested reader can follow additional links listed in the Appendix.

*4.2.1 General Synoptic Surveys in the Visible Regime*

Many current wide-field transient surveys are targeted toward the discovery and characterization of GRB afterglows. One example is he *Robotic Optical Transient Search Experiment* (ROTSE-III; Akerlof et al. 2003; http://www.rotse.net), uses 4 telescopes with a FOV = 3.4 deg$^2$ to survey 1260 deg$^2$ to $R$~17.5 mag (e,g, Rykoff et al. 2005; Yost et al. 2007). ROTSE searches for astrophysical optical transients on time scales of a fraction of a second to a few hours. While the primary incentive of this experiment has been to find the GRB afterglows, additional variability studies have been enabled, e.g., searches for RR Lyrae.

The *Palomar-Quest* (PQ; Djorgovski et al. 2008; http://palquest.org) survey was a collaborative project between groups at Yale University and Caltech, with an extended network of collaborations with other groups world-wide. The data were obtained at the Palomar Observatory's Samuel Oschin telescope (the 48-inch Schmidt) using the QUEST-2, 112-CCD, 161 Mpix camera (Baltay et al. 2007). Approx. 45% of the telescope time was used for the PQ survey, with the rest being used by the NEAT survey for near-Earth asteroids, and miscellaneous other projects. The survey started in the late summer of 2003, and ended in September 2008.

In the first phase (2003-2006), data were obtained in the drift scan mode in 4.6° wide strips of a constant Dec, in the range −25° < Dec < +25°, excluding the Galactic plane. The total area coverage is ~ 15,000 deg$^2$, with multiple passes, typically 5 – 10, but up to 25, with time baselines ranging from hours to years. Typical area coverage rate was up to ~ 500 deg$^2$/night in 4 filters. The raw data rate was on average ~ 70 GB per clear night. About 25 TB of usable data have been collected in the drift scan mode. Two filter sets were used, Johnson *UBRI* and SDSS *griz*. Effective exposures are ~ 150/(cos δ) sec per pass, with limiting magnitudes are $r$ ~ 21.5, $i$ ~ 20.5, $z$ ~ 19.5, $R$ ~ 22, and $I$ ~ 21 mag, depending on the seeing, lunar phase, etc. Photometric calibrations were done by matching to the SDSS. In the second phase of the survey (2007-2008), data were obtained mostly in the traditional point-and-track mode, in a single, wide-band red filter (RG610). The coverage and the cadence were largely optimized for the nearby supernova search, in collaboration with the LBNL NSNF group, and a search for dwarf planets, in collaboration with M. Brown at Caltech.

Data were processed with several different pipelines, optimized for different scientific goals: the Yale pipeline (Andrews et al. 2008), designed for a search for gravitationally lensed quasars; the Caltech real-time pipeline, used for real-time detections of transient events, and the LBNL NSNF pipeline, based on image subtraction and designed for detection of nearby SNe. The data are being publicly released. Coadded images (the *DeepSky* project, Nugent et al., in prep.; http://supernova.lbl.gov/~nugent/deepsky.html) reach to R ~ 23 mag over ~ 20,000 deg$^2$.

The *Nearby Supernova Factory* (NSNF; http://snfactory.lbl.gov; Aldering et al. 2002), operated by the Lawrence Berkeley National Laboratory (LBNL) searches for type Ia SNe in the redshift range 0.03 < $z$ < 0.08, in order to establish the low-redshift anchor of the SN Ia Hubble diagram. This experiment ran from 2003 to 2008 using data obtained by the NEAT and PQ surveys, resulting in several hundred spectroscopically confirmed SNe.

The *Palomar Transient Factory* (PTF; http://www.astro.caltech.edu/ptf; Rau et al. 2009, Law et al. 2009) operates in the way that is very similar to the PQ survey, on the same telescope, but with a much better camera that was previously used on the CFHT. The data are taken in a point-and-stare mode, with 2 exposures per field in a given night, mostly in the broad *R* and *G* bands. PTF reaches a comparable depth to its predecessor, PQ, and covers a few hundred deg$^2$/night.

The overall area coverage is ~ ½ of the entire sky. The survey mostly uses a cadence optimized for a SN discovery. The principal data processing pipeline is an updated version of the LBNL NSNF image subtraction pipeline. The survey has been very productive, but the data are proprietary to the PTF consortium, at least as of this writing.

The *Catalina Real-Time Transient Survey* (CRTS; Drake et al. 2009, 2012; Djorgovski et al. 2011a, Mahabal et al. 2011; http://crts.caltech.edu) leverages existing synoptic telescopes and image data resources from the Catalina Sky Survey (CSS) for near-Earth objects and potential planetary hazard asteroids (NEO/PHA), conducted by Edward Beshore, Steve Larson, and their collaborators at the Univ. of Arizona. The Solar System objects remain the domain of the CSS survey, while CRTS aims to detect astrophysical transient and variable objects, using the same data stream. Real-time processing, characterization, and distribution of these events, as well as follow-up observations of selected events constitute the CRTS.

CRTS utilizes three wide-field telescopes: the 0.68-m Schmidt at Catalina Station, AZ (CSS), the 0.5-m Uppsala Schmidt (Siding Spring Survey, SSS) at Siding Spring Observatory, NSW, Australia, and the Mt. Lemmon Survey (MLS), a 1.5-m reflector located on Mt. Lemmon, AZ. Each telescope employs a camera with a single, cooled, 4096×4096 pixel, back-illuminated, unfiltered CCD. They are operated for 23 nights per lunation, centered on new moon. Most of the observable sky is covered up to 4 times per lunation. The total area coverage is ~ 30,000 deg$^2$, as it excludes the Galactic plane within $|b| < 10° - 15°$. In a given night, 4 images of the same field are taken, separated by ~ 10 min, for a total time baseline of ~30 min between first and last images. The combined data streams cover up to 2,500 deg$^2$/night to a limiting magnitude of V ~ 19 – 20 mag, and up to 275 deg$^2$/night to V ~ 21.5 mag. On time scales of a few weeks, the combined survey covers most of the available sky from Arizona and Australia, with a few sets of exposures.

Optical transients (OTs) and highly variable objects (HVOs) are detected as sources that display significant changes in brightness, as compared to the baseline comparison images and catalogs with significantly fainter limiting magnitudes; these are obtaining by meadian stacking of > 20 previous observations, and reach ~ 2 mag deeper. The search is performed in the catalog domain, but an image subtraction pipeline is also being developed. Data cover time baselines from 10 min to several years.

CRTS practices an *open data* philosophy: detected transients are published electronically in real time, using a variety of mechanisms, so that they can be followed up by anyone. Archival light curves for ~ 500 million objects, and all of the images are also being made public.

The *All Sky Automated Survey* (ASAS; http://www.astrouw.edu.pl/asas/; Pojmanski 1997) covers the entire sky using a set of small telescopes (telephoto lenses) at Las Campanas Observatory in Chile, and Haleakala, Maui, in *V* and *I* bands. The limiting magnitude are $V \sim 14$ mag and $I \sim 13$ mag, with ~ 14 arcsec/pixel sampling . The *ASAS-3 Photometric V-band Catalogue* contains over 15 million light curves; a derived product is the on-line *ASAS Catalog of Variable Stars* (ACVS), with ~ 50 thousand selected objects.

### 4.2.2 *Supernova Surveys*

Supernova (SN) searches have been providing science drivers for a number of surveys. We recall that the original Zwicky survey from Palomar was motivated by the need to find more SNe for the follow-up studies. In addition to those described in Sec. 4.2.1, some of the dedicated SN surveys include:

The *Calan/Tololo Supernova Search* (Hamuy et al. 1993) operated from 1989 to 1993, and provided some of the key foundations for the subsequent use of type Ia SNe as standard candle, which eventually led to the evidence for dark energy.

The *Supernova Cosmology Project* (SCP, Perlmutter et al. 1998, 1999) was designed to measure the expansion of the universe using high redshift type-Ia SNe as standard candles. Simultaneously, the *High-z Supernova Team* (Filippenko & Riess 1998) was engaged in the same pursuit. While the initial SCP results were inconclusive (Perlmutter et al. 1997), in 1998 both experiments discovered that the expansions of the universe was accelerating (Perlmutter et al. 1998, Riess et al. 1998), thus providing a key piece of evidence for the existence of the dark energy (DE). These discoveries are documented and discussed very extensively in the literature.

The *Equation of State: SupErNovae trace Cosmic Expansion* project (ESSENCE; http://www.ctio.noao.edu/essence) was a SN survey optimized to constrain the equation of state (EoS) of the DE, using type Ia SNe at redshifts $0.2 < z < 0.8$. The goal of the project was to determine the value of the EoS parameter w to within 10%. The survey used 30 half-nights per year with the CTIO Blanco 4 m telescope and the MOSAIC camera between 2003 and 2007 and discovered 228 type-Ia SNe. Observations were made in the *VRI* passbands with a cadence of a few days. SNe were discovered using an image subtraction pipeline, with candidates inspected and ranked for follow-up by eye.

Similarly, the *Supernova Legacy Survey* (SNLS; http://www.cfht.hawaii.edu/SNLS) was designed to precisely measure several hundred type Ia SNe at redshifts $0.3 < z < 1$ or so (Astier et al. 2006). Imaging was done at the Canada-France-Hawaii Telescope (CFHT) with the *Megaprime* 340 Megapixel camera, providing a 1 deg$^2$ FOV, in the course of 450 nights, between 2003 and 2008. About 1,000 SNe were discovered and followed up. Spectroscopic follow-up was dome using the 8-10 m class telescopes, resulting in ~ 500 confirmed SNe.

The *Lick Observatory Supernova Search* (LOSS; Filippenko et al. 2001; http://astro.berkeley.edu/bait/public_html/kait.html) is and ongoing search for low-z (z < 0.1) SNe that began in 1998. The experiment makes use of the 0.76 m Katzman Automatic Imaging Telescope (KAIT). Observations are taken with a 500×500 pixel unfiltered CCD, with an FoV of 6.7×6.7 arcmin$^2$, reaching R ~ 19 mag in 25 seconds (Li et al. 1999). As of 2011, the survey made 2.3 million observations, and discovered a total of 874 SNe by repeatedly monitoring ~ 15,000 large nearby galaxies (Leaman et al. 2011). The project is designed to tie high-z SNe to their more easily observed local counterparts, and also to rapidly responded to Gamma-Ray Burst (GRB) triggers and monitor their light curves.

The *SDSS-II Supernova Search* was designed to detect type Ia SNe at intermediate redshifts ($0.05 < z < 0.4$). This experiment ran 3-month campaigns from September to November during 2005 to 2007, by doing repeated scans of the SDSS Stripe 82 (Sako et al. 2008). Detection were made using image subtraction, and candidate type-Ia SNe were selected using the SDSS 5-bandpass imaging. The experiment resulted in the discovery and spectroscopic confirmation of ~ 500 type Ia SNe, and 80 core-collapse SNe. Although the survey was designed to discover SNe, the data have been made public, enabling the study of numerous types of variables objects including variables star and AGN (Sesar et al. 2007; Bhatti et al. 2010).

Many other surveys and experiments have contributed to this vibrant field. Many amateur astronomers have also contributed; particularly noteworthy is the Puckett Observatory World Supernova Search (http://www.cometwatch.com/search.html) that involves several telescopes, and B. Monard's Supernova search.

### 4.2.3 Synoptic Surveys for Minor Bodies in the Solar System

The discovery that NEO events were potentially hazardous to human life motivated programs to discover the extent and nature of these objects, in particular the largest ones which could yield a significant destruction, the Potentially Hazardous Asteroids (PHAs). A number of NEO surveys began working with NASA in order to fulfill the congressional mandate to discover all NEOs larger than 1 km in diameter. The number of NEOs discovered each year continues to increase, as surveys increase in sensitivity, with the current rate of discovery of ~ 1,000 NEOs per year.

The search for NEOs (Near Earth Objects) with CCDs largely began with the discovery of the asteroid 1989 UP by the *Spacewatch* survey (http://spacewatch.lpl.arizona.edu). They also made the first automated NEO detection, of the asteroid 1990 SS. The survey uses the Steward Observatory 0.9 m telescope, and a camera covering 2.9 deg2. A 1.8 m telescope with a FOV or 0.8 deg2 was commissioned in 2001. Until 1997, Spacewatch was the major discoverer of NEOs, in addition to a number of Trans-Neptunian objects (TNO).

The *Near-Earth Asteroid Tracking* (NEAT; http://neat.jpl.nasa.gov) began in 1996 and ran until 2006. Initially, the survey used a three-CCD camera mounted on the Samuel Oschin 48-inch Schmidt telescope on Palomar Mountain, until it was replaced by the PQ survey's 160 Megapixel camera in 2003. In addition, in 2000 the project began using data from the Maui Space Surveillance Site 1.2 m telescope. A considerable number of NEOs was found by this survey. The work was combined with Mike Brown's search for dwarf planets, and resulted in a number of such discoveries, also causing the demotion of Pluto to a dwarf planet status. The data were also used by the PQ survey team for a SN search.

The *Lincoln Near-Earth Asteroid Research* survey, conducted by the MIT Lincoln Laboratory (LINEAR; http://www.ll.mit.edu/mission/space/linear) began in 1997, using four 1 m telescopes at the White Sands Missile Range in New Mexico (Viggh et al. 1997). The survey has discovered ~ 2,400 NEOs to date, among 230,000 other asteroids and comet discoveries.

The *Lowell Observatory Near-Earth-Object Search* (LONEOS; http://asteroid.lowell.edu/asteroid/loneos/loneos.html) ran from 1993 to 2008. The survey used a 0.6 m Schmidt telescope with an 8.3 deg FOV, located near Flagstaff, Arizona. It discovered 289 NEOs.

The *Catalina Sky Survey* (CSS; http://www.lpl.arizona.edu/css) began in 1999 and was extensively upgraded in 2003. It uses the telescopes and produces the data stream that is also used by the CRTS survey, described above. It became the most successful NEO discovery survey in 2004, and it has so far discovered ~ 70% of all known NEOs. On 6 October 2008 a small NEO, 2008 TC3, was the discovered by their 1.5 m telescope. The object was predicted to impact the Earth within 20 hours of discovery. The asteroid disintegrated in the upper atmosphere over the Sudan desert; 600 fragments were recovered on the ground, amounting to "an extremely low cost sample return mission". This was the first time that an asteroid impact on Earth has been accurately predicted.

The Pan-STARRS survey, discussed above, also has a significant NEO search component. Another interesting concept was proposed by Tonry (2011).

Recently, the *NEOWISE* program (a part of the WISE mission) discovered 130 NEOs, among a much larger number of asteroids.

One major lesson of the PQ survey was that a joint asteroid/transient analysis is necessary, since slowly moving asteroids represented a major "contaminant" in a search for astrophysical transients. A major lesson of the Catalina surveys is that the same data streams can be shared very effectively between the NEO and transient surveys, thus greatly increasing the utility of the shared data.

*4.2.4 Microlensing Surveys*

The proposition of detecting and measuring dark matter in in the form of *Massive Compact Halo Objects* (MACHOs) using gravitational microlensing was first proposed by Paczynski (1986). The advent of large format CCDs made the measurements of millions of stars required to detect gravitation lensing a possibility in the crowded stellar fields toward the Galactic Bulge and the Magellanic Clouds. Three surveys began searching for gravitation microlensing and the MACHO and EROS (*Exp'erience pour la Recherche d'Objets Sombres*) groups simultaneously announced the discovery of events toward the LMC in 1993 (Alcock et al. 1993; Aubourg et al. 1993). Meanwhile, the *Optical Gravitational Lensing Experiment* (OGLE) collaboration searched and found the first microlensing toward the Galactic bulge (Udalski et al. 1993). All three surveys continued to monitor ~ 100 deg$^2$ in fields toward the Galactic Bulge, and discovered hundreds of microlensing events. A similar area was covered toward the Large Magellanic Cloud, and a dozen events were discovered. This result limited the contribution of MACHO to halo dark matter to 20% (Alcock et al. 2000).

It was also predicted that the signal of planets could be detected during microlensing events (Mao and Paczynski 1991). Searches for such events were undertaken by the *Global Microlensing Alert Network* (GMAN) and PLANET (*Probing Lensing Anomalies NETwork*) collaborations beginning in 1995. The first detection of planetary lensing was in 2003, when the microlensing event OGLE-2003-BLG-235 was found to harbor a Jovian mass planet (Bond et al. 2004). Udalski et al. (2002) used the OGLE-III camera and telescope to perform a dedicated search for transits and discovered the first large sample of transiting planets.

Additional planetary microlensing event surveys such as *Microlensing Follow-Up Network* (MicroFun) and *RoboNet-I,II* (Mottram & Fraser 2008, Tsapras et al. 2009) continue to find planets by following microlensing events discovered by OGLE-IV and MOA (*Microlensing Observations in Astrophysics*; Yock et al. 2000) surveys, which discover hundreds of microlensing events toward Galactic Bulge each year. Additional searches to further quantify the amount of matter in MACHOs have been undertaken in microlensing have been carried out by AGAPE (Ansari et al. 1997) toward M31 and the LMC by the SuperMACHO project (http://www.ctio.noao.edu/supermacho/lightechos/SM/sm.html).

In addition to the microlensing events, the surveys revealed the presence of tens of thousands of variables stars and others sources, as also predicted by Paczynski (1986). The data collected by these surveys represent a gold mine of information on variable stars of all kinds.

*4.2.5 Radio Synoptic Surveys*

There has been a considerable interest recently in the exploration of the time domain in the radio regime. Ofek et al. (2011) reviewed the transient survey literature in the radio regime. The new generation of radio surveys and facilities is enabled largely by the advances in the radio electronics (themselves driven by the mobile computing and cell phone demands), as well as the ICT. These are reflected in better receivers and correlators, among other elements, and since much of this technology is off the shelf, the relatively low cost of individual units enables production of large new radio facilities that collect vast quantities of data. Some of them include:

The *Allen Telescope Array* (ATA; http://www.seti.org/ata) consists of 42 6-m dishes covering a frequency range of 0.5 – 11 GHz. It has a maximum baseline of 300m and a field-of-view of 2.5 deg at 1.4 GHz. A number of surveys have been conducted with the ATA. The *Fly's Eye Transient* search (Siemion et al. 2012) was a 500-hour survey over 198 square degrees at 209 MHz with all dishes pointing in different directions, looking for powerful millisecond timescale transients. The *ATA Twenty-centimeter Survey* (ATATS; Croft et al. 2010, 2011) was a survey at 1.4 GHz constructed from 12 epoch visits over an area of ~700 deg$^2$ for rare bright transients and to prove the wide-field capabilities of the ATA. Its catalog contains 4984 sources above 20 mJy (>90% complete to ~40 mJy) with positional accuracies better than 20 arcsec and is comparable to the NVSS. Finally the *Pi GHz Sky Survey* (PiGSS; Bower et al. 2010, 2011) is a 3.1 GHz radio continuum survey of ~250,000 radio sources in the 10,000 deg$^2$ region of sky with b > 30° down to ~1 mJy, with each source being observed multiple times. A subregion of ~11 deg$^2$ will also be repeatedly observed to characterize variability on timescales of days to years.

*The Low Frequency Radio Array* (LOFAR; http://www.lofar.org/) comprises 48 stations, each consisting of two types of dipole antenna: 48/96 LBA covering a frequency range of 10 – 90 MHz and 48/96 4x4 HBA tiles covering 110 – 240 MHz. There is a dense core of 6 stations, 34 more distributed throughout the Netherlands and 8 elsewhere in Europe (UK, France, Germany). LOFAR can simultaneously operate up to 244 independent beams and observe sources at Dec > -30. A recent observation reached 100 μJy (L-band) over a region of ~60 deg$^2$ comprising ~10,000 sources. A number of surveys are planned including all-sky surveys at 15, 30, 60, and 120 MHz and ~1000 deg$^2$ at 200 MHz to study star, galaxy and large-scale structure formation in the early universe, intercluster magnetic fields and explore new regions of parameter space for serendipitous discovery, probe the epoch of reionization and zenith and Galactic Plane transient monitoring programs at 30 and 120 MHz.

The field is largely driven towards the next generation radio facility, the Square Kilometer Array (SKA), described below. Two prototype facilities are currently being developed:

One of the SKA precursors is the *Australian Square Kilometer Array Pathfinder* (ASKAP; http://www.atnf.csiro.au/projects/askap), comprising of 36, 12 m antennae covering a frequency range of 700 – 1800 MHz. It has a FOV of 30° and a maximum baseline of 6 km. In 1 hr it will be able to reach 30 μJy per beam with a resolution of 7.5 arcsec. An initial construction of 6 antennae (BETA) is due for completion in 2012. The primary science goals of ASKAP include galaxy formation in the nearby universe through extragalactic HI surveys (WALLABY, DINGO), formation and evolution of galaxies across cosmic time with high-resolution, confusion-limited continuum surveys (EMU, FLASH), characterization of the radio transient sky through detection and monitoring of transient and variable sources (CRAFT, VAST) and the evolution of magnetic fields in galaxies over cosmic time through polarization surveys (POSSUM).

*MeerKAT* (http://www.ska.ac.za/meerkat/index.php) is the South African SKA pathfinder due for completion in 2016. It consists of 64, 13.5m dishes operating at ~1 GHz and 8 – 14.5 GHz with a baseline between 29 m to 20 km. It complements ASKAP with a larger frequency range and greater sensitivity but a smaller field of view. It has enhanced surface brightness sensitivity with its shorter and longer baselines and will also have some degree of phased element array capability. The primary science drivers cover the same type of SKA precursor science area as ASKAP, but MeerKAT will focus on those areas where its has unique capabilities - these include extremely sensitive studies of neutral hydrogen in emission to $z \sim 1.4$ and highly sensitive continuum surveys to µJy levels at frequencies as low as 580 MHz. An initial test bed of seven dishes (KAT-7) have been constructed and is now being used as an engineering and science prototype.

### 4.2.6 Other Wavelength Regimes

Given that the variability is common to most point sources at high energies, and the fact that the detectors in this regime tend to be photon counting with a good time resolution, surveys in that regime are almost by definition synoptic in character.

An excellent example of a modern high-energy survey is the *Fermi Gamma-Ray Space Telescope* (FGST; http://www.nasa.gov/fermi), launched in 2008, that is continuously monitoring the γ-ray sky. Its Large Area Telescope (LAT) is a pair-production telescope that detects individual photons, with a peak effective area of ~ 8,000 $cm^2$ and a field of view of ~ 2.4 sr. The observations can be binned into different time resolutions. The stable response of LAT and a combination of deep and fairly uniform exposure produces an excellent all-sky survey in the 100 MeV to 100 GeV range to study different types of γ-ray sources. This has resulted in a series of ongoing source catalog releases (Abdo et al. 2009, 2010, and more to come). The current source count approaches 2,000, with ~ 60% of them identified on other wavelengths.

### 4.3 Towards the Petascale Data Streams and Beyond

The next generation of sky surveys will be largely synoptic in nature, and will move us firmly to the Petascale regime. We are already entering it with the facilities like PanSTARRS, ASKAP, and LOFAR. These new surveys and instruments not only depend critically on the ICT, but also push it to a new performance regime.

The *Large Synoptic Sky Survey* (LSST; Tyson et al. 2002, Ivezic et al. 2009; http://lsst.org) is a wide-field telescope that will be located at Cerro Paranal in Chile. The primary mirror will be 8.4 m in diameter, but because of the 3.4 m secondary, the collecting area is equivalent to that of a ~ 6.7 m telescope. While its development is still in the relatively early stages as of this writing, the project has a very strong community support, reflected in its top ranking in the 2010 Astronomy and Astrophysics Decadal Survey produced by the U.S. National Academy of Sciences, and it may become operational before this decade is out. The LSST is planned to produce a 6-bandpass (0.3 – 1.1 micron) wide-field, deep astronomical survey of over 20,000 $deg^2$ of the Southern sky, with many epochs per field. The camera will have a ~ 3.2 Gigapixel detector array covering ~ 9.6 $deg^2$ in individual exposures, with 0.2 arcsec pixels.

LSST will take more than 800 panoramic images each night, with 2 exposures per field, covering the accessible sky twice each week. The data (images, catalogs, alerts) will be continuously generated and updated every observing night. In addition, calibration and co-added images, and the resulting catalogs, will be generated on a slower cadence, and used for data quality assessments. The final source catalog is expected to have more than 20 billion rows, comprising 30 TB of data per night, for a total of 60 PB over the envisioned duration of the survey. Its scientific goals and strategies are described in detail in the *LSST Science Book* (Ivezic et al. 2009). Processing and analysis of this huge data stream poses a number of challenges in the arena of real-time data processing, distribution, archiving, and analysis.

The currently ongoing renaissance in the continuum radio astronomy at cm and m scale wavelengths is leading towards the facilities that will surpass even the data rates expected for the LSST, and that will move us to the Exascale regime.

The *Square Kilometer Array* (SKA; http://skatelescope.org) will be the world's largest radio telescope, hoped to be operational in the mid-2020's. It is envisioned to have a total collecting area of approximately 1 million $m^2$ (thus the name). It will provide continuous frequency coverage from 70 MHz to 30 GHz employing phased arrays of dipole antennas (low frequency), tiles (mid frequency) and dishes (high frequency) arranged over a region extending out to ~ 3,000 km. Its key science projects will focus on studying pulsars as extreme tests of general relativity, mapping a billion galaxies to the highest redshifts by their 21-cm emission to understand galaxy formation and evolution and dark matter, observing the primordial distribution of gas to probe the epoch of reionization, investigating cosmic magnetism and surveying all types of transient radio phenomena, and so on.

The data processing for the SKA poses significant challenges, even if we extrapolate Moore's law to its projected operations. The data will stream from the detectors into the correlator at a rate of ~4.2 PB/s, and then from the correlator to the visibility processors at rates between 1 and 500 TB/s, depending on the observing mode, which will require processing capabilities of ~200 Pflops – 2.5 Eflops. Subsequent image formation needs ~10 Pflops to create data products (~ 0.5 – 10 PB/day), which would be available for science analysis and archiving, the total computational costs of which could easily exceed those of the pipeline. Of course, this is not just a matter of hardware provision, even if it is special purpose built, but also high computational complexity algorithms for wide field imaging techniques, deconvolution, Bayesian source finding, and other tasks. Each operation will also place different constraints on the computational infrastructure, with some being memory bound and some CPU bound that will need to be optimally balanced for maximum throughput. Finally the power required for all this processing will also need to be addressed – assuming the current trends, the SKA data processing will consume energy at a rate of ~1 GW. These are highly non-trivial hardware and infrastructure challenges.

The real job of science, data analysis and knowledge discovery, starts after all this processing and delivery of processed data to the archives. Effective, scalable software and methodology needed for these tasks does not exist yet, at least in the public domain.

### *4.4 Deep Field Surveys*

A counterpart to the relatively shallow, wide-field surveys discussed above are various deep fields that cover much smaller, selected areas, generally based on an initial deep set of observations by one of the major space-based observatories. Their scientific goals are almost always in the arena of galaxy formation and evolution: the depth is required in order to obtain adequate measurements of very distant ones, and the areas surveyed are too small to use for much of the Galactic structure work. This is a huge and vibrant field of research, and we cannot do it justice in the limited space here; we describe very briefly a few of the more popular deep surveys. More can be found, e.g., in the reviews by Ferguson et al. (2000), Bowyer et al. (2000), Brandt & Hassinger (2005), and the volume edited by Cristiani et al. (2002).

The prototype of these is the *Hubble Deep Field* (HDF; Williams et al. 1996; http://www.stsci.edu/ftp/science/hdf/hdf.html). It was imaged with the HST in 1995 using the Director's Discretionary Time, with 150 consecutive HST orbits and the WFPC2 instrument, in four filters ranging from near-UV to near-IR, F300W, F450W, F606W, and F814W. The ~ 2.5 arcmin field is located at RA = 12:36:49.4, Dec = +62:12:58 (J2000); a corresponding HDF-South, centered at RA = 22:32:59.2, Dec = –60:33:02.7 (J2000) was observed in 1998. An even deeper *Hubble Ultra Deep Field* (HUDF; Beckwith et al. 2006), centered at RA = 03:32:39.0, Dec = –27:47:29 (J2000) and covering 11 arcmin$^2$, was observed with 400 orbits in 2003-2004, using the ACS instrument in 4 bands: F435W, F606W, F775W, and F850LP, and with 192 orbits in 2009 with the WFC3 instrument in 3 near-IR filters, F105W, F125W, and F160W. Depths equivalent to V ~ 30 mag were reached. There has been a large number of follow-up studies of these deep fields, involving many ground-based and space-based observatories, mainly on the subjects of galaxy formation and evolution at high redshifts.

The *Groth Survey Strip* field, is a 127 arcmin$^2$ region that has been observed with the HST in both broad V and I bands, and a variety of ground-based and space-based observatories. It was expanded to the *Extended Groth Strip* in 2004-2005, covering 700 arcmin$^2$ using 500 exposures with the ACS instrument on the HST. The *All-Wavelength Extended Groth Strip International Survey* (AEGIS; Davis et al. 2007) continues to study this at multiple wavelengths.

The *Chandra Deep Field South* (CDF-S; http://www.eso.org/~vmainier/cdfs_pub/; Giacconi et al. 2001) was obtained by the *Chandra* X-ray telescope looking at the same patch of the sky for 11 consecutive days in 1999-2000, for a total of 1 million seconds. The field covers 0.11 *deg*$^2$ centered at RA = 03:32:28.0, Dec = –27:48:30 (J2000). These X-ray observations were followed up extensively by many other observatories.

The *Great Observatories Origins Deep Survey* (GOODS; http://www.stsci.edu/science/goods/; Dickinson et al. 2003, Giavalisco et al. 2004) builds on the HDF-N and CDF-S by targeting fields centered on those areas, and covers approximately 0.09 *deg*$^2$ using a number of NASA and ESA space observatories: *Spitzer, Hubble, Chandra, Herschel, XMM-Newton*, etc., as well as a number of deep ground-based studies and redshift surveys. The main goal is to study the distant universe to faintest fluxes across the electromagnetic spectrum, with a focus on the formation and evolution of galaxies.

The *Subaru Deep Field* (http://www.naoj.org/; Kashikawa et al. 2004) was observed initially over 30 nights at the 8.2 m Subaru telescope on Mauna Kea, using the *SupremeCam* instrument. The field, centered at RA = 13:24:38.9, Dec = +27:29:26 (J2000), cover a patch of 34 by 27 arcmin$^2$. It was imaged in the *BVRi'z'* bands, and narrow bands centered at 8150 Å and 9196 Å. One of the main aims was to catalog Lyman-break galaxies out to large redshifts, and get samples of Ly$\alpha$ emitters (LAE) as probes of the very early galaxy evolution. Over 200,000 galaxies were detected, yielding samples of hundreds of high-redshift galaxies. These were followed by extensive spectroscopy and imaging on other wavelengths.

The *Cosmological Evolution Survey* (COSMOS; http://cosmos.astro.caltech.edu; Scoville et al. 2007) was initiated as an HST Treasury Project, but it expanded to include deep observation from a range of facilities, both ground-based and space-based. The 2 $deg^2$ field is centered at RA = 10:00:28.6, Dec = +02:12:21.0 (J2000). The HST observations were carried out in 2004-2005. These were followed by a large number of ground-based and space-based observatories. The main goals were to detect 2 million objects down to the limiting magnitude $I_{AB} > 27$ mag, including > 35,000 Lyman Break Galaxies and extremely red galaxies out to $z \sim 5$. The follow-up included an extensive spectroscopic and imaging on other wavelengths

*The Cosmic Assembly Near-IR Deep Extragalactic Legacy Survey* (CANDELS; 2010-2013; http://candels.ucolick.org/; Koekemoer et al. 2011) uses deep HST imaging of over 250,000 galaxies with WFC3/IR and ACS instruments to study galactic evolution at $1.5 < z < 8$. The surveys is at 2 depths: moderate (2 orbits) over 0.2 $deg^2$, and deep (12 orbits) over 0.04 $deg^2$. An additional goal is to refine the constraints on time variation of cosmic-equation of state parameter leading to a better understanding of the dark energy.

A few ground-based deep surveys (in addition to the follow-up of those listed above) are also worthy of note. They include:

The *Deep Lens Survey* (DLS; 2001-2006; Wittman et al. 2002; http://dls.physics.ucdavis.edu) covered ~ 20 $deg^2$ spread over 5 fields, using CCD mosaics at KPNO and CTIO 4 m telescopes, with 0.257 arcsec/pixel, in *BVRz'* bands. The survey also had a significant synoptic component (Becker et al. 2004).

The *NOAO Deep Wide Field Survey* (NDWFS; http://www.noao.edu/noao/noaodeep; Jannuzi & Dey 1999) covered ~ 18 $deg^2$ spread over 2 fields, also using CCD mosaics at KPNO and CTIO 4 m telescopes, in $B_wRI$ bands, reaching point source limiting magnitudes of 26.6, 25.8, and 25.5 mag, respectively. These fields have been followed up extensively at other wavelengths.

CFHT Legacy Survey (CFHTLS; 2003-2009; http://www.cfht.hawaii.edu/Science/CFHLS) is a community-oriented service project that covered 4 $deg^2$ spread over 4 fields, using the MegaCam imager at the CFHT 3.6 m telescope, over an equivalent of 450 nights, in *u\*g'r'i'z'* bands. The survey served a number of different projects.

Deep fields have had a transformative effect on the studies of galaxy formation and evolution. They also illustrated a power of multi-wavelength studies, as observations in each regime leveraged each other. Finally, they also illustrated a power of the combination of deep space-based imaging, followed by deep spectroscopy with large ground-based telescopes.

Many additional deep or semi-deep fields have been covered from the ground in a similar manner. We cannot do justice to them or to the entire subject in the limited space available here, and we direct the reader to a very extensive literature that resulted from these studies.

*4.5 Spectroscopic Surveys*

While imaging tends to be the first step in many astronomical ventures, physical understanding often comes from spectroscopy. Thus, spectroscopic surveys naturally follow the imaging ones. However, they tend to be much more focused and limited scientifically: there are seldom surprising new uses for the spectra, beyond the original scientific motivation.

Two massive, wide-field redshift surveys that have dramatically changed the field used highly multiplexed (multi-fiber or multi-slit) spectrographs; they are the 2dF and SDSS. A few other significant previous redshift surveys were mentioned in Sec. 2.

Several surveys were done using the 2-degree field (2dF) multi-fiber spectrograph at the Anglo-Australian Telescope (AAT) at Siding Spring, Australia. The *2dF Galaxy Redshift Survey* (2dFGRS; 1997-2002; Colless 1999; Colless et al. 2001) observed ~ 250,000 galaxies down to $m_B \approx 19.5$ mag. The *2dF Quasar Redshift Survey* (2QZ; 1997-2002; Croom et al. 2004) observed ~ 25,000 quasars down to $m_B \approx 21$ mag. The spectroscopic component of the SDSS eventually produced redshifts and other spectroscopic measurements for ~ 930,000 galaxies and ~ 120,000 quasars, and nearly 500,000 stars and other objects. These massive spectroscopic surveys greatly expanded our knowledge of the LSS, the evolution of quasars, and led to numerous other studies.

Deep imaging surveys tend to be followed by the deep spectroscopic ones, since redshifts are essential for their interpretation and scientific uses. Some of the notable examples include:

The *Deep Extragalactic Evolution Probe* (DEEP; Vogt et al. 2005, Davis et al. 2003) survey is a two-phase project using the 10 m Keck telescopes to obtain spectra of faint galaxies over ~ 3.5 deg$^2$, to study the evolution of their properties and evolution of clustering out to $z \sim 1.5$. Most of the data were obtained using the DEIMOS multi-slit spectrograph, with targets preselected using photometric redshifts. Spectra were obtained for ~ 50,000 galaxies down to ~ 24 mag, in the redshift range $z \sim 0.7 - 1.55$ with candidates, with ~ 80% yielding redshifts.

The *VIRMOS-VLT Deep Survey* (VVDS; Le Fevre et al. 2005) is an ongoing comprehensive imaging and redshift survey of the deep universe, complementary to the Keck DEEP survey. An area of ~16 deg$^2$ in four separate fields is covered using the VIRMOS multi-object spectrograph on the 8.2 m VLT telescopes. With 10 arcsec slits, spectra can be measured for 600 objects simultaneously. Targets were selected from a *UBRI* photometric survey (limiting magnitude $I_{AB}$ = 25.3 mag) carried out with the CFHT. A total of over 150,000 redshifts will be obtained (~ 100,000 to $I_{AB}$ = 22.5 mag, ~ 50,000 to $I_{AB}$ = 24 mag, and ~ 1,000 to $I_{AB}$ = 26 mag), providing insight into galaxy and structure formation over a very broad redshift range, $0 < z < 5$.

zCOSMOS (Lilly et al. 2007) is a similar survey on the VLT using VIRMOS, but only covering the 1.7 deg$^2$ COSMOS ACS field down to a magnitude limit of $I_{AB} < 22.5$ mag. Approximately 20,000 galaxies were measured over the redshift range $0.1 < z < 1.2$, comparable in survey parameters to the 2dFGRS, with a further 10,000 within the central 1 deg$^2$, selected with photometric redshifts in the range $1.4 < z < 3.0$.

The currently ongoing *Baryon Oscillation Spectroscopic Survey* (BOSS), due to be completed in early 2014 as part of SDSS-III, will cover an area of ~ 10,000 deg$^2$ and map the spatial distribution of ~ 1.5 million luminous red galaxies to $z = 0.7$, and absorption lines in the spectra of ~ 160,000 quasars to $z \sim 3.5$. Using the characteristic scale imprinted by the baryon acoustic oscillation in the early universe in these distributions as a standard ruler, it will be able to determine the angular diameter distance and the cosmic expansion rate with a precision of ~ 1 to 2%, in order to constrain theoretical models of dark energy.

Its successor, the *BigBOSS* survey (http://bigboss.lbl.gov) will utilize a purpose-built 5,000 fiber mutiobject spectrograph, to be mounted at the prime focus of the KPNO 4 m Mayall telescope, covering a 3° FOV. The project will conduct a series of redshift surveys over ~ 500 nights spread over 5 years, with a primary goal of constraining models of dark energy, using different observational tracers: clustering of galaxies out to $z \sim 1.7$, and Ly$\alpha$ forest lines in the spectra of quasars at $2.2 < z < 3.5$. The survey plans to obtain spectra of $\sim 20$ million galaxies and $\sim 600,000$ quasars over a 14,000 deg$^2$ area, in order to reach these goals.

Another interesting, incipient surveys is the *Large Sky Area Multi-Object Fiber Spectroscopic Telescope* (LAMOST; http://www.lamost.org/website/en), one of the National Major Scientific Projects undertaken by the Chinese Academy of Science. A custom-built 4 m telescopes has a 5° FOV, accommodating 4,000 optical fibers for a highly multiplexed spectroscopy.

A different approach uses panoramic, slitless spectroscopy surveys, where a dispersing element (an objective prism, grating, or a grism – grating ruled on a prism) is placed in the front of the telescope optics, thus providing wavelength-dispersed images for all objects. This approach tends to work only for the bright sources in ground-based observations, since any given pixel gets the signal form the object only at the corresponding wavelength, but the sky background that dominates the noise has contributions from all wavelengths. For space-based observations, this is much less of a problem. Crowding can also be a problem, due to the overlap of adjacent spectra. Traditionally, slitless spectroscopy surveys have been used to search for emission-line objects (e.g., star-forming or active galaxies and AGN), or objects with a particular continuum signature (e.g., low metallicity stars).

One good example is the set of surveys done at the Hamburg Observatory, covering both the Northern sky from the Calar Alto observatory in Spain, and the Southern sky from ESO (http://www.hs.uni-hamburg.de/EN/For/Exg/Sur), that used a wide-angle objective prism survey to look for AGN candidates, hot stars and optical counterparts to ROSAT X-ray sources. For each $5.5 \times 5.5$ deg$^2$ field in the survey area, two objective prism plates were taken, as well as an unfiltered direct plate to determine accurate positions and recognize overlaps. All plates were subsequently digitized and, under good seeing conditions, a spectral resolution of 45A at H$\gamma$ was be achieved. Candidates selected from these slitless spectra were followed by targeted spectroscopy, producing $\sim 500$ AGN and over 2,000 other emission line sources, and a large number of extremely metal-poor stars.

## *4.6 Figures of Merit*

Since sky surveys cover so many dimensions of the OPS, but only a subset of them, comparisons of surveys can be difficult if not downright meaningless. What really matters is a scientific discovery potential, which very much depends on what are the primary science goals; for example, a survey optimized to find supernovae may not be very efficient for studies of galaxy clusters, and vice versa. However, for just about any scientific application, area coverage and depth, and, in the case of synoptic sky surveys also the number of epochs per field, are the most basic characteristics, and can be compared fairly, at least in any given wavelength regime. Here we attempt to define some general indicators of the scientific discovery potential for sky surveys, based on these general parameters.

Often, but misleadingly, surveys are compared using the *etendue*, the product of the telescope area, $A$, and the solid angle subtended by the individual exposures, $\Omega$. However, $A\Omega$ simply reflects the properties of a telescope and the instrument optics. It implies nothing whatsoever about a survey that may be done using that telescope and instrument combination, as it does not say anything about the depth of the individual exposures, the total area coverage, and the number of exposures per field. The $A\Omega$ is the same for a single, short exposure, and for a survey that has thousands of fields, several bandpasses, deep exposures, and hundreds of epochs for each field. Clearly, a more appropriate figure of merit (FoM), or a set thereof, is needed.

We propose that a fair measure of a general scientific potential of a survey would be a product of a measure of its average depth, and its coverage of the relevant dimensions of the OPS. Or: how deep, how wide, and how often, and for how long.

As a quantitative measure of depth in an average single observation, we can define a quantity that is roughly proportional to the S/N for background-limited observations for an unresolved source, namely:

$$D = [ A \times t_{exp} \times \varepsilon ]^{1/2} / FWHM$$

where $A$ is the effective collecting area of the telescope in m$^2$, $t_{exp}$ is the typical exposure length in sec, $\varepsilon$ is the overall throughput efficiency of the telescope+instrument, and $FWHM$ is the typical PSF or beam size full-width at half-maximum (i.e., the seeing, for the ground-based visible and NIR observations) in arcsec.

The coverage of the OPS depends on the number of bandpasses (or, in the case of radio or high energy surveys, octaves; or, in the case of spectroscopic surveys, the number of independent resolution elements in a typical survey spectrum), $N_b$, and the total survey area covered, $\Omega_{tot}$, expressed as a fraction of the entire sky. In the case of synoptic sky surveys, additional relevant parameters include the area coverage rate regardless of the bandpass (in deg$^2$/night, or as a fraction of the entire sky per night), $R = d\Omega/dt$, the number of exposures per field per night regardless of the bandpass, $N_e$, or, for the survey as a whole, the average total number of visits per field in a given bandpass, $N_{avg}$. For a single-epoch imaging survey, $N_{avg} = 1$. The total number of all exposures for a given field, regardless of the bandpass, is $N_{tot} = N_b \times N_{avg}$. The coverage along the time axis of the OPS is roughly proportional to $N_{tot}$.

Thus, we define the Scientific Discovery Potential (*SDP*) FoM as:

$$SDP = D \times \Omega_{tot} \times N_b \times N_{avg}$$

It probably makes little sense to compare single-pass and synoptic surveys using this metric: single-pass surveys are meaningfully compared mainly by the depth, the area coverage, and the number of bandpasses, whereas for the synoptic sky surveys, the number of passes per field matters a lot, since their focus is on the exploration of the time domain.

These FoM pertain to the survey as a whole, and are certainly applicable for an archival research. If what matters is a discovery *rate* of transient events *(TDR)*, we define another FoM:

$$TDR = D \times R \times N_e$$

That is, the area covered in a given night is observed $N_e$ times, in any bandpass. Note: *TDR* is not the actual number of transients per night, as that depends on many other factors, but it should be roughly proportional to it. Obviously, the longer one runs the survey, the more transient events are found.

Both of these indicators are meaningful for the imaging surveys. For a spectroscopic (targeted) survey, a useful FoM may be a product of the depth parameter, $D$, and the number of targets covered.

These FoM are in some sense the bare minimum of information needed to characterize or compare surveys. They do not account for things like the sky background and transparency, the total numbers of sources detected (which clearly depends strongly on the Galactic latitude), the width of the bandpasses (wavelength resolution), the dynamical range of the data, the quality of the calibration, instrumental noise and artifacts, the angular resolution, the uniformity of coverage as a function of both position on the sky and time, and, in the case of synoptic sky surveys, the cadences. These FoM also do not account for the operational parameters such as the data availability and access, the time delay between the observations and the event publishing, etc., all of which affect the science produced by the surveys.

They also implicitly assume that surveys in the same wavelength regime and with the same type of data, e.g., direct imaging, are being compared. A comparison of surveys at different wavelengths makes sense only in some specific scientific context. For example, large collecting areas of radio telescopes do not make them that much more scientifically effective than, say, X-ray telescopes, with their small effective collecting areas. A fair comparison of surveys with different types of data, e.g., images and spectra, would be even more difficult and context dependent.

Another approach is to compare the hyper-volumes covered by the surveys in some subset of the OPS. For example, we can define the 4-dimensional Survey Hyper-Volume (*SHV*) as:

$$SHV = \Omega_{tot} \times \log(f_{max}/f_{min}) \times (\lambda_{max}/\lambda_{min}) \times N_{tot}$$

where $\Omega_{tot}$ is the total sky area covered, now expressed as a fraction of the entire sky; $(f_{max}/f_{min})$ is the dynamical range of flux measurements, with $f_{min}$ being the limiting flux for significant detections and $f_{max}$ the saturation level, $(\lambda_{max}/\lambda_{min})$ is the dynamical range of wavelengths covered, and $N_{tot}$ is the mean number of exposures per field; $N_{tot} = N_b$ for the single-pass surveys. Thus, we are carving out a hyper-volume along the spatial, flux, wavelength, and possibly also temporal axes. A further refinement would be to divide each of the terms in this product by the resolution of measurements on the corresponding axis. That would give a number of the independent *SHV* elements in the survey.

The *SHV* as defined above provides a FoM that *can* be used, at least in principle, to compare surveys on different wavelengths, and even with different types of data (images, spectra, time series, etc.). However, it does not provide much of the other, potentially relevant information described above, and it does not reflect the *depth* of the survey. Thus the SHV favors smaller telescopes.

All of these FoM provide a generality at the cost of a scientific specificity. They are more appropriate for surveys with a broad range of scientific goals, than for the narrowly focused ones. Like any tool, they should be used with a caution.

Table 2 gives our best estimates of these FoM for a number of selected surveys. It is meant to be illustrative rather than definitive. The survey parameters used are based on the information available in the literature and/or the Web, and sometimes our informed estimates.

An alternative, probably far too simple way of comparing surveys, is by their quantity of data collected, e.g., in TB. We do not advocate such simplemindedness, since not all bits are of an equal value, but many important properties of surveys, including their scientific potential, *do* correlate at least roughly with the size of the data sets.

If the main scientific goals are statistical studies that require large samples of objects, then a reasonable figure of merit may be the product of the number of detected sources, $N_{src}$, and the independent measured parameters (attributes) per source, $N_{param}$. The later is also a measure of the complexity of the data set. For a spectroscopic survey, simply the number of spectra obtained is a reasonable indicator of the survey's scope.

| *Survey* | *A* $[m^2]$ | $t_{exp}$ [s] | $\varepsilon$ | *FWHM* arcsec | *D* | $N_b$ | $N_{avg}$ | $\Omega_{tot}$ $[deg^2]$ (frac) | *SDP* | $N_e$ | *R* $deg^2/nt$ (frac) | *TDR* | log $f_{max}/f_{min}$ | $\lambda_{max}/\lambda_{min}$ | *SHV* |
|---|---|---|---|---|---|---|---|---|---|---|---|---|---|---|---|
| SDSS (imaging) | 4.0 | 54 | 0.4 | 1.5 | 6.2 | 5 | 1 | 14500 (0.35) | 10.85 | 1 | … | … | 4.2 | 3 | 22 |
| 2MASS | 1.3 | 45 | 0.35 | 2.5 | 1.8 | 3 | 1 | 41250 (1.0) | 5.4 | 1 | … | … | 4.5 | 2 | 27 |
| UKIDSS | 11.3 | 40 | 0.35 | 0.8 | 15.7 | 4 | 1 | 7200 (0.175) | 11.0 | 1 | … | … | 4.5 | 2.5 | 7.9 |
| NVSS | 12,200 | 30 | 0.7 | 45 | 11.2 | 1 | 1 | 34000 (0.82) | 9.2 | 1 | … | … | 5 | 1.1 | 4.5 |
| FIRST | 12,200 | 165 | 0.7 | 5 | 237.4 | 1 | 1 | 9900 (0.24) | 57 | 1 | … | … | 5 | 1.1 | 1.3 |
| SDSS stripe 82 | 4.0 | 54 | 0.4 | 1.5 | 6.2 | 5 | 300 | 300 (0.007) | 67 | 1 | … | … | 4.2 | 3 | 137 |
| PanSTARRS PS1, 3 yrs | 2.5 | 30 | 0.5 | 1.0 | 6.1 | 6 | 12 | 30000 (0.73) | 320 | 1 | 6000 (0.145) | 0.887 | 4.2 | 3 | 550 |
| PQ/NEAT | 1.0 | 150 | 0.4 | 2 | 3.9 | 4,1 ⟨2⟩ | 50 | 20000 (0.48) | 190 | 2 | 500 (0.012) | 0.095 | 4.0 | 2.5 | 480 |
| CRTS 4 yr | 2.32 | 30 | 0.4 | 3 | 1.8 | 1 | 400 | 34000 (0.82) | 590 | 4 | 2000 (0.048) | 0.345 | 4.0 | 2.5 | 3280 |
| PTF 3 yr | 1.0 | 60 | 0.4 | 1.8 | 2.7 | 2 | 50 | 22000 (0.53) | 145 | 2 | 1000 (0.024) | 0.131 | 4.5 | 1.5 | 360 |
| SkyMapper 3yr (est.) | 1.1 | 110 | 0.5 | 2 | 3.3 | 6 | 36 | 22000 (0.53) | 380 | 2 | 800 (0.019) | 0.128 | 4.2 | 3 | 1440 |
| LSST 10 yr (est.) | 35.3 | 15 | 0.5 | 0.8 | 20.3 | 5 | 2000 | 25000 (0.60) | 122000 | 2 | 4000 (0.097) | 3.94 | 4.2 | 2.5 | 63000 |

**Table 2.** Parameters and figures of merit for some of the wide-field surveys. The parameters listed are our best estimates on the basis of the published information. See the text for more details and websites. For PQ/NEAT, some data were taken in 4 bandpasses, and some in 1; we use 2 as an effective average. The PTF takes ~ 70% of the data in the *R* band, and ~ 30% in the *g* band; we averaged the numbers. For CRTS, the three sub-surveys were combined, and we use the effective duration of 4 years, as already accomplished as of this writing. The PTF and the PS1 are currently in progress; the numbers for the SkyMapper and the LSST are the projections for the future. We assumed the quoted durations of 3 years for the PS1 survey, the PTF, and the SkyMapper survey, and 10 years for the LSST. These durations scale directly with the assumed $N_{tot}$, and thus affect the SDP and the SHV figures of merit. For a more fair comparison, the ongoing and the proposed synoptic surveys should probably be compared on a per-year basis, an exercise we leave to the reader.

## 5. From the Raw Data to Science-Ready Archives

Surveys, being highly data-intensive ventures where uniformity of data products is very important, pose a number of data processing and analysis challenges (Djorgovski & Brunner 2001). Broadly speaking, the steps along the way include: obtaining the data (telescope, observatory and instrument control software), on-site data processing, if any, detection of discrete sources and measurements of their parameters in imaging surveys, or extraction of 1-dimensional spectra and measurements of the pertinent features in the spectroscopic ones, data calibrations, archiving and dissemination of the results, and finally the scientific analysis and exploration. Typically, there is a hierarchy of ever more distilled and value-added data products, starting from the raw instrument output and ending with ever more sophisticated descriptions of detected objects.

The great diversity of astronomical instruments and types of data with their specific processing requirements are addressed elsewhere in these volumes. Likewise, the data archives, virtual observatory and astroinformatics issues, data mining and the problem-specific scientific analysis are beyond the scope of this review. Here we address the intermediate steps that are particularly relevant for the processing and dissemination of survey data.

Many relevant papers for this subject can be found in the Astronomical Data Analysis and Software Systems (ADASS) and Astronomical Data Analysis (ADA) conference series, and in the SPIE volumes that cover astronomical instruments and software. Another useful reference is the volume edited by Graham, Fitzpatrick, & McGlynn (2008).

### *5.1 Data Processing Pipelines*

The actual gathering and processing of the raw survey data encompasses many steps, which can be often performed using a dedicated software pipeline that is usually optimized for the particular instrument, and for the desired data output; that by itself may introduce some built-in biases, but if the original raw data are kept, they can always be reprocessed with improved or alternative pipelines.

Increasingly, we see surveys testing their pipelines extensively with simulated data, well before the actual hardware is built. This may reflect a cultural influence of the high-energy physics, as hey are increasingly participating in the major survey projects, and data simulations are essential in their field. However, one cannot simulate the problems that are discovered only when the real data are flowing.

The first step involves hardware-specific data acquisition software, used to operate the telescopes and the instruments themselves. In principle, this is not very different from the general astronomical software used for such purposes, except that the sky surveying generally requires a larger data throughput, a very stable and reliable operation over long stretches of time, and considerably greater data flows than is the case for most astronomical observing. In most cases, additional flux calibration data are taken, possibly with separate instruments or at different times. Due to the long amounts of time required to complete a survey (often several years), a great deal of care must be exercised to monitor the overall performance of the survey in order to ensure a uniform data quality.

Once the raw images, spectra, data cubes, or time series are converted in a form that has the instrumental signatures removed, and the data are represented as a linear intensity as a function of the spatial coordinates, wavelength, time, etc., the process of source detection and characterization starts. This requires a good understanding of the instrumental noise properties, which determines some kind of a detection significance threshold: one wants to go as deep as possible, but not count the noise peaks. In other words, we always try to maximize the completeness (the fraction of the real sources detected) while minimizing the contamination (the fraction of the noise peaks mistaken for real sources). In the linear regime of a given detector, the former should be as close to unity, and the latter as close to zero as possible. Both deteriorate at the fainter flux levels as the S/N drops. Typically, a detection limit is taken as a flux level where the completeness falls below 90% or so, and contamination increases above 10% or so. However, *significant* detections actually occur at some higher flux level.

Most source detection algorithms require a certain minimum number of adjacent or connected pixels above some signal-to-noise thresholds for detection. The optimal choice of these thresholds depends on the power spectrum of the noise. In many cases, the detection process involves some type of smoothing or optimal filtering, e.g., with a Gaussian whose width approximates that of an unresolved point source. Unfortunately, this also builds in a preferred scale for source detection, usually optimized for the unresolved sources (e.g., stars) or the barely-resolved ones (e.g., faint galaxies), which are the majority. This is a practical solution, but with the obvious selection biases, with the detection of sources depending not only on their flux, but also on their shape or contrast: there is almost always a *limiting surface brightness* (averaged over some specific angular scale) in addition to the *limiting flux*. A surface brightness bias is always present at some level, whether it is actually important or not for a given scientific goal. Novel approaches to source, or, more accurately, structure detection involve so-called multi-scale techniques (e.g., Aragon-Calvo et al. 2007).

Once individual sources are detected, a number of photometric and structural parameters are measured for them, including fluxes in a range of apertures, various diameters, radial moments of the light distribution, etc., from which a suitably defined, intensity-weighted centroid is computed. In most cases, the sky background intensity level is determined locally, e.g., in a large aperture surrounding each source; crowding and contamination by other nearby sources can present problems and create detection and measurement biases. Another difficult problem is deblending or splitting of adjacent sources, typically defined as a number of distinct, adjacent intensity peaks connected above the detection surface brightness threshold. A proper approach keeps track of the hierarchy of split objects, usually called the parent object (the blended composite), the children objects (the first level splits), and so on. Dividing the total flux between them and assigning other structural parameters to them are nontrivial issues, and depend on the nature of the data and the intended scientific applications.

Object detection and parameter measurement modules in survey processing systems often use (or are based on) some standard astronomical program intended for such applications, e.g., *FOCAS* (Jarvis & Tyson 1981), *SExtractor* (Bertin & Arnouts 1996), or *DAOPHOT* (Stetson 1987), to mention just a few popular ones. Such programs are well documented in the literature. Many surveys have adopted modified versions of these programs, optimized for their own data and scientific goals.

Even if custom software is developed for these tasks, the technical issues are very similar. It is generally true that all such systems are built with certain assumptions about the properties of sources to be detected and measured, and optimized for a particular purpose, e.g., detection of faint galaxies, or accurate stellar photometry. Such data may serve most users well, but there is always a possibility that a custom reprocessing for a given scientific purpose may be needed.

At this point (or further down the line) astrometric and flux calibrations are applied to the data, using the measured source positions and instrumental fluxes. Most surveys are designed so that improved calibrations can be reapplied at any stage. In some cases, it is better to apply such calibration after the object classification (see below), as the transformations may be different for the unresolved and the resolved sources. Once the astrometric solutions are applied, catalogs from adjacent or overlapping survey images can be stitched together.

In the mid-1990's, the rise of the TB-scale surveys brought the necessity of dedicated, optimized, highly automated pipelines, and databases to store, organize and access the data. One example is DPOSS, initially processed using *SKICAT* (Weir et al. 1995c), a system that incorporated databases and machine learning, which was still a novelty at that time. SDSS developed a set of pipelines for the processing and cataloguing of images, their astrometric and photometric calibration, and for the processing of spectra; additional, specialized pipelines were added later, to respond to particular scientific needs. For more details, see, e.g., York et al. (2000), Lupton et al. (2001), Stoughton et al. (2002), and the documentation available at the SDSS website.

A major innovation of SDSS (at least for the ground-based data; NASA missions and data centers were also pioneering such practices in astronomy) was the effective use of databases for data archiving and the Web-based interfaces for the data access, and in particular the *SkyServer* (Szalay, Gray, et al. 2001, 2002). Multiple public data releases were made using this approach, with the last one (DR8) in 2011, covering the extensions (SDSS-II and SDSS-III) of the original survey. By the early 2000's, similar practices were established as standard for most other surveys; for example, the UKIDSS data processing is described by Dye et al. (2006) and Hodgkin et al. (2009).

Synoptic sky survey added the requirement of data processing in real time, or as close to it as possible, so that transient events can be identified and followed in a timely fashion. For example, the PQ survey (2003-2008) had 3 independent pipelines, a traditional one at Yale University (Andrews et al. 2008), an image subtraction pipeline optimized for SN discovery at the LBNL *Nearby Supernova Factory* (Aldering et al. 2002), and the *Palomar-Quest Event Factory* (Djorgovski et al. 2008) pipeline at Caltech (2005-2008), optimized for a real-time discovery of transient events. The latter served as a basis for the CRTS survey pipeline (Drake et al. 2009). Following in the footsteps of the PQ survey, the PTF survey operates in a very similar manner, with the updated version of the NSNF near-real-time image subtraction pipeline for discovery of transients, and a non-time-critical pipeline for additional processing at IPAC.

An additional requirement for the synoptic sky surveys is a timely and efficient dissemination of transient events, now accomplished through a variety of electronic publishing mechanisms. Perhaps the first modern example was the *Gamma-Ray Coordinates Network* (GCN; Barthelmy et al. 2000; http://gcn.gsfc.nasa.gov), that played a key role in cracking the puzzle of the GRBs. For the ground-based surveys, the key effort was the *VOEventNet* (VOEN; Williams & Seaman 2006; http://voeventnet.caltech.edu), that developed the *VOEvent*, now an adopted standard protocol for astronomical event electronic publishing and communication, and deployed it in an experimental robotic telescope network with a feedback, using the PQEF as a primary testbed. This effort currently continues through the *SkyAlert* facility (http://skyalert.org; Williams et al. 2009), that uses the CRTS survey as its primary testbed. A variety of specific event dissemination mechanism have been deployed, using the standard webpages, RSS feeds, and even the mobile computing and social media.

*5.2 Source and Event Classification*

Object classification, e.g., as stars or galaxies in the visible and near-IR surveys, but more generally as resolved and unresolved sources, is one of the key issues. Classification of objects is an important aspect of characterizing the astrophysical content of a given sky survey, and for many scientific applications one wants either stars (i.e., unresolved objects) or galaxies; consider for example studies of the Galactic structure and studies of the large-scale structure in the universe. More detailed morphological classification, e.g., Hubble types of detected galaxies, may be also performed if the data contain sufficient discriminating information to enable it. Given the large data volumes involved in digital sky surveys, object classification must be automated, and in order to make it really useful, it has to be as reliable and objective as possible, and homogeneous over the entire survey. Often, the *classification limit* is more relevant than the detection limit, for definition of statistical samples of sources (e.g., stars, galaxies, quasars).

In most cases, object classification is based on some quantitative measurements of the image morphology for the detected sources. For example, star-galaxy separation in optical and near-IR surveys uses the fact that all stars (and also quasars) would be unresolved point sources, and that the observed shape of the light distribution would be given by the point-spread function, whereas galaxies would be more extended. This may be quantified through various measures of the object radial shape or concentration, e.g., moments of the light distribution in various combinations. The problem of star-galaxy separation thus becomes a problem of defining a boundary in some parameter space of observed object properties, which would divide the two classes. In simplest approaches such a dividing line or surface is set empirically, but more sophisticated techniques use artificial intelligence methods, such as the Artificial Neural Nets or Decision Trees (e.g., Weir et al. 1995a, Odewahn et al. 2004, Ball et al. 2006, Donalek et al. 2008). They require a training data set of objects for which the classification is known accurately from some independent observations. Because of this additional information input, such techniques can outperform the methods where survey data alone are used to decide on the correct object classifications.

There are several practical problems in this task. First, fainter galaxies are smaller in angular extent, thus approaching stars in their appearance. At the fainter flux levels the measurements are noisier, and thus the two types of objects become indistinguishable. This sets a classification limit to most optical and near-IR surveys, which is typically at a flux level a few times higher than the detection limit. Second, the shape of the point-spread function may vary over the span of the survey, e.g., due to the inevitable seeing variations. This may be partly overcome by defining the point-spread function locally, and normalizing the structural parameters of objects so that the unresolved sources are the same over the entire survey. In other words, one must define the unresolved source template that would be true locally, but may (and usually does) vary globally. Furthermore, this has to be done automatically and reliably over the entire survey data domain, which may be very heterogeneous in depth and intrinsic resolution. Additional problems include object blending, saturation of signal at bright flux levels, detector nonlinearities, etc., all of which modify the source morphology, and thus affect the classification.

The net result is that the automated object classification process is always stochastic in nature. Classification accuracies better than 90% are usually required, but accuracies higher than about 95% are generally hard to achieve, especially at faint flux levels.

In other situations, e.g., where the angular resolution of the data is poor, or where nonthermal processes are dominant generators of the observed flux, morphology of the objects may have little meaning, and other approaches are necessary. Flux ratios in different bandpasses, i.e., the spectrum shape, may be useful in separating different physical classes of objects.

A much more challenging task is the automated classification of transient events discovered in synoptic sky surveys (Djorgovski et al. 2006, 2011b, Mahabal et al. 2005, 2008ab, Bloom et al. 2012). Physical classification of the transient sources is the key to their interpretation and scientific uses, and in many cases scientific returns come from the follow-up observations that depend on scarce or costly resources (e.g., observing time at larger telescopes). Since the transients change rapidly, a rapid (as close to the real time as possible) classification, prioritization, and follow-up are essential, the time scale depending on the nature of the source, which is initially unknown. In some cases the initial classification may remove the rapid-response requirement, but even an archival (i.e., not time-critical) classification of transients poses some interesting challenges.

This entails some special challenges beyond traditional automated classification approaches, which are usually done in some feature vector space, with an abundance of self-contained data derived from homogeneous measurements. Here, the input information is generally sparse and heterogeneous: there are only a few initial measurements, and the types differ from case to case, and the values have differing variances; the contextual information is often essential, and yet difficult to capture and incorporate in the classification process; many sources of noise, instrumental glitches, etc., can masquerade as transient events in the data stream; new, heterogeneous data arrive, and the classification must be iterated dynamically. Requiring a high completeness, a low contamination, and the need to complete the classification process and make an optimal decision about expending valuable follow-up resources (e.g., obtain additional measurements using a more powerful instrument at a certain cost) in real time are challenges that require some novel approaches.

The first challenge is to associate classification probabilities that any given event belongs to a variety of known classes of variable astrophysical objects and to update such classifications as more data come in, until a scientifically justified convergence is reached. Perhaps an even more interesting possibility is that a given transient represents a previously unknown class of objects or phenomena, that may register as having a low probability of belonging to any of the known data models. The process has to be *as automated as possible, robust, and reliable*; it has to operate from *sparse and heterogeneous data*; it has to maintain a *high completeness* (not miss any interesting events) yet a *low false alarm rate*; and it has to *learn* from the past experience for an ever improving, evolving performance.

Much of the initial information that may be used for the event classification is archival, implying a need for a good VO-style infrastructure. Much of the relevant information is also contextual: for example, the light curve and observed properties of a transient might be consistent with both it being a cataclysmic variable star, a blazar, or a supernova. If it is subsequently known that there is a galaxy in close proximity, the supernova interpretation becomes much more plausible. Such information, however, can be characterized by high uncertainty and absence, and by a rich structure – if there were two candidate host galaxies, their morphologies, distances, and luminosities become important, e.g., is this type of supernova more consistent with being in the extended halo of a large spiral galaxy or in close proximity to a faint dwarf galaxy? The ability to incorporate such contextual information in a quantifiable fashion is essential. There is a need to find a means of harvesting the human pattern recognition skills, especially in the context of capturing the relevant contextual information, and turning them into machine-processible algorithms.

These challenges are still very much a subject of an ongoing research. Some of the relevant papers and reviews include Mahabal et al. (2010abc), Djorgovski et al. (2011b), Richards et al. (2011), and Bloom & Richards (2012), among others.

*5.3 Data Archives, Analysis and Exploration*

In general, the data processing flow is from the pixel (image) domain to the catalog domain (detected sources with measured parameters). This usually results in a reduction of the data volume by about an order of magnitude (this factor varies considerably, depending on the survey or the data set), since most pixels do not contain statistically significant signal from resolved sources. However, the ability to store large amounts of digital image information on-line opens up interesting new possibilities, whereby one may want to go back to the pixels and remeasure fluxes or other parameters, on the basis of the catalog information. For example, if a source was detected (i.e., cataloged) in one bandpass, but not in another, it is worth checking if a marginal detection is present even if it did not make it past the statistical significance cut the first time; even the absence of flux is sometimes useful information.

Once all of the data has been extracted from the image pixels by the survey pipeline software, it must be stored in some accessible way in order to facilitate scientific exploration. Simple user file systems and directories are not suitable for really large data volumes produced by sky surveys. The transition to Terascale data sets in the 1990's necessitated use of dedicated database software. Using a database system provides significant advantages (e.g., powerful and complex query expressions) combined with a rapid data access. Fortunately, commercially available database systems can be adopted for astronomical uses. Relational databases accessed using the *Structured Query Language* (SQL) tend to dominate at this time, but different architectures may be better scaleable for the much larger data volumes in the future.

A good example of a survey archive is the *SkyServer* (Szalay, Gray, et al. 2001, 2002; http://skyserver.org/), that provides access to data (photometry and spectra) for objects detected in the different SDSS data sets. This supports more than just positional searching – it offers the ability to pose arbitrary queries (expressed in SQL) against the data so that, for example, one can find all merging galaxy pairs or quasars with a broad absorption line and a nearby galaxy within 10 arcsec. Users can get their own work areas so that query results can be saved and files uploaded to use in queries (as user-supplied tables) against the SDSS data.

Currently, most significant surveys are stored in archives that are accessible through the Internet, using a variety of web service interfaces. Their interoperability is established through the Virtual Observatory framework. Enabling access to such survey archives via web services and not just web pages means that programs can be written to automatically analyze and explore vast amounts of data. Whole pipelines can be launched to coordinate and federate multiple queries against different archives, potentially taking hundreds of hours to automatically find the rarest species of objects. Of course, the utility of any such archive is only as good as the metadata provided and the hardest task is often figuring out exactly how the same concept is represented in different archives, for example, one archive might report flux in a particular passband and another magnitude, and manually reconciling these.

The *Semantic Web* is an emerging technology that can help solve these challenges (Antoniou & van Harmelen 2004). It is based on machine-processable descriptions of concepts, and it goes beyond simple term matching with expressions of concept hierarchies, properties and relationships allowing knowledge discovery. It is a way of encoding a domain expertise (e.g., in astronomy) in a way that may be used by a machine. Ultimately, it may lead to data inferencing by artificial intelligence (AI) engines. For example, discovering that a transient detection has no previous outburst history, is near a galaxy and has a spectrum with silicon absorption but no hydrogen, a system could reason that it is likely to be a Type Ia supernova and therefore its progenitor was a white dwarf and so perform an appropriate archival search to find it.

*Cloud computing* is an emerging paradigm that may well change the ways we approach data persistence, access, and exploration. Commodity computing brings economies of scale, and it effectively outsources a number of tedious tasks that characterize data-intensive science. It is possible that in the future most of our data, survey archives included, and data mining and exploration services for knowledge discovery, will reside in the Cloud.

Most of the modern survey data sets are so information-rich, that a wide variety of different scientific studies can be done with the same data. Therein lies their scientific potential (Djorgovski et al. 1997b, 2001abc, 2002, Babu & Djorgovski 2004, and many others). However, this requires some powerful, general tools for the exploration, visualization, and analysis of large survey data sets. Reviewing them is beyond the scope of this chapter, but one recent example is the *Data Mining and Exploration* system (DAME; Brescia et al. 2010, 2011; http://dame.dsf.unina.it); see also the review by Ball & Brunner (2010). The newly emerging discipline of Astroinformatics may provide a research framework and environment that would foster development of such tools.

## 6. Concluding Comments

In this chapter we have attempted to summarize, albeit briefly, the history and the state of the art of sky surveys. However, this is a very rapidly evolving field, and the reader is advised to examine the subsequent literature for the updates and descriptions of new surveys and their science.

In some fields, surveys completely dominate the observational approaches; for example, cosmology, either as a quest to describe the global properties of the universe, the nature of the dark energy, etc., or the history of structure and galaxy formation and evolution, is now tackled largely through large surveys, both from ground and space. Surveys discover cosmic explosions, extrasolar planets, and even new or predicted phenomena.

Sky surveys have transformed the ways in which astronomy is done, and pushed it from the relative data poverty to a regime of an immense data overabundance. They are the by far the largest generators of data in astronomy, and they have already enabled a lot of important science, and will undoubtedly continue to do so. They have also fostered the emergence of the Virtual Observatory framework and Astroinformatics as means of addressing both the challenges and the opportunities brought by the exponential data growth. They also represent a superb starting point for education and public outreach, e.g., with the *Google Sky* and the *WorldWide Telescope* (WWT; http://www.worldwidetelescope.org) sky browsers.

Surveys have also revitalized the role of small telescopes in the era of giant ones, both for the surveying itself, and for the immediate imaging and photometric follow-up (Djorgovski 2002).

Small telescopes do not imply a small science. Survey-based astronomy is inherently systemic, requiring a full hierarchy of observational facilities, since much of the survey-based science is in the follow-up studies of selected sources. Mutual leveraging of survey and follow-up telescopes, on-line archives and cyber-infrastructure, creates an added value for all of them.

There is, however, one significant bottleneck that we can already anticipate in the survey-driven science: the follow-up spectroscopy of interesting sources selected from imaging surveys. While there seems to be a vigorous ongoing and planned activity to map and monitor the sky in many ways and many wavelengths, spectroscopic surveys will be necessary in order to interpret and understand the likely overabundance of potentially interesting objects. This looming crisis may seriously limit the scientific returns from the ongoing and future surveys.

Another important lesson is that the cost of these data-intensive projects is increasingly dominated by the cost of software development, implementation, and maintenance. Nobody has ever *under*estimated the cost of software. Our community has to develop more effective ways of sharing and leveraging software efforts. This remains as one of the key motivations behind the VO and Astroinformatics.

In addition to their roles as scientific and technological catalysts, surveys have also started to change the sociology and culture of astronomy, by opening new modes of research, new kinds of problems to be addressed, requiring new skills, and new modes of scientific publishing and knowledge preservation. This cultural shift is both inevitable and profoundly transformational. Other sciences have undergone comparable or greater changes, driven by the ways in which problems are defined, and data are obtained and analyzed; biology is a good example.

Some sociological changes may be a mixed blessing. By their nature, surveys tend to require large teams, since that can help secure the necessary resources (funding, observing time) and the manpower. Many astronomers are uneasy about this trend towards the high-energy physics mode of research. Generating large data sets requires large-scale efforts. However, important discoveries are still being made at all scales, from individuals and small groups, to large collaborations. Proposed survey science tends to be a committee-designed science, and thus often safe, but unimaginative; actual survey science tends to be dominated by the unexpected uses of the data and surprises.

One important way in which surveys have changed astronomy is their role as an intermediary step between the sky and the scientist. The large information content of modern sky surveys enables numerous studies that go well beyond the original purposes. The traditional approach where we observe selected targets and make new discoveries using such primary observational data is still with us, and will remain. However, there is now a new way of observing the sky, through its representation in the survey archives, using software instruments. It is now possible to make significant observational advances and discoveries without ever going to a telescope. Thus we see a rise in prominence of archival research, which can be as cutting-edge as any observations with the world's largest telescopes and space observatories.

This type of research requires new computational science skills, from data farming (databases, their interoperability, web services, etc.) to data mining and knowledge discovery. The methods and the tools that work efficiently in the Megabyte to Gigabyte regime usually do not scale to Terabytes and Petabytes of data, let alone the greatly increased complexity of the modern data sets. Effective visualization tools and techniques for high-dimensionality parameter sets are another critical issue. We need new kinds of expertise for the new, data-rich and data-intensive astronomy in the 21$^{st}$ century. As the science evolves, so does its methodology: we need both new kinds of tools, and the people who know how to use them.

Unfortunately, we are currently neither training properly the new generations of researchers in these requisite skills, nor rewarding the career paths that bridge astronomy and ICT. The culture of academia changes slowly, and these educational and professional recognition issues may be among the key obstacles in our path towards the full scientific utilization of the great and growing data abundance brought by the modern sky surveys.

Astronomy is not alone among the sciences facing these challenges. Interdisciplinary exchanges in the context of e-Science, cyber-infrastructure, and science informatics can help us tackle these important issues more efficiently. All of them signal a growing virtualization of science, as most of our work moves into the cyberspace.

To end on a positive note, we are likely entering a new golden age of discovery in astronomy, enabled by the exponential growth of the ICT, and the resulting exponential growth of data rates, volumes, and complexity. Any science must rest on the data as its empirical basis, and sky surveys are increasingly playing a fundamental role in this regard in astronomy. We have really just started to exploit them, and the future will bring many new challenges and opportunities for discovery.

**Acknowledgments:**

We are indebted to many colleagues and collaborators over the years, especially the key members of the survey teams: Nick Weir, Usama Fayyad, Joe Roden, Reinaldo de Carvalho, Steve Odewahn, Roy Gal, Robert Brunner, and Julia Kennefick in the case of DPOSS; Eilat Glikman, Roy Williams, Charlie Baltay, David Rabinowitz, and the rest of the Yale team in the case of PQ; and Steve Larson, Ed Beshore, and the rest of the Arizona and Australia team in the case of CRTS. Likewise, we acknowledge numerous additional colleagues and collaborators in the Virtual Observatory and the Astroinformatics community, especially Alex Szalay, Jim Gray, Giuseppe Longo, Yan Xu, Tom Prince, Mark Stalzer, and many others. Several tens of excellent undergraduate research students at Caltech contributed to our work through the years, many of them supported by the Caltech's SURF program. And last, but not least, the staff of Palomar, Keck, and other observatories, who helped the data flow. Our work on sky surveys and their exploration has been supported in part by the NSF grants AST-0122449, AST-0326524, AST-0407448, CNS-0540369, AST-0834235, AST-0909182 and IIS-1118041; the NASA grant 08-AISR08-0085; and by the Ajax and Fishbein Family Foundations. Some of the figures in this paper have been produced using an immersive VR visualization software, supported in part by the NSF grant HCC-0917814. We thank numerous colleagues, and in particular H. Bond, G. Longo, M. Strauss, and M. Kurtz, whose critical reading improved the text. Finally, we thank The Editors for their saintly patience while waiting for the completion of this chapter.

**Appendix: A Partial List of Sky Surveys, Facilities, and Archives as of 2011**

We provide this listing as a handy starting point for a further exploration. The listing is probably biased and subjective, and is certainly incomplete, and we apologize for any omissions. The websites given are the best available as of 2011, and a judicious use of search engines can be used to provide updates and corrections. Likewise, VO-related services can be an effective way to discover available data resources for a particular problem.

An exhaustive listing of surveys is maintained by the *NASA/IPAC Extragalactic Database* (NED), currently available at http://ned.ipac.caltech.edu/samples/NEDmdb.html

2MASS = Two Micron All Sky Survey, http://www.ipac.caltech.edu/2mass
ADS = Astrophysics Data System, http://adswww.harvard.edu/
AGAPE = Andromeda Galaxy Amplified Pixels Experiment,
 http://www.ing.iac.es/PR/SH/SH2006/agape.html
AGILE = Astro-rivelatore Gamma a Immagini LEggero, http://agile.rm.iasf.cnr.it/publ02.html
Akari, http://www.ir.isas.jaxa.jp/ASTRO-F/Outreach/index_e.html
ALFALFA = Arecibo Legacy Fast ALFA Survey, http://egg.astro.cornell.edu
ALMA = Atacama Large Millimeter/sub-millimeter Array, http://www.eso.org/public/teles-
 instr/alma.html
APASS = The AAVSO Photometric All-Sky Survey, http://www.aavso.org/apass
ASAS-1,2,3 = All Sky Automated Survey, http://www.astrouw.edu.pl/asas/
ASKAP = Australian Square Kilometer Array Pathfinder,
 http://www.atnf.csiro.au/projects/askap
ATA = Allen Telescope Array, http://www.seti.org/ata
CANDELS = Cosmic Assembly Near-infrared Deep Extragalactic Legacy Survey,
 http://candels.ucolick.org
CARMA = Combined Array for Research in Millimeter-wave Astronomy,
 http://www.mmarray.org/
CXC = Chandra X-Ray Center, http://cxc.harvard.edu/
CDFS = Chandra Deep Field South, http://www.eso.org/~vmainier/cdfs_pub/
CDS = Centre de Données astronomiques de Strasbourg, http://cdsweb.u-strasbg.fr/
COBE = Cosmic Background Explorer, http://lambda.gsfc.nasa.gov/product/cobe
CoRoT = COnvection ROtation and planetary Transits, http://smsc.cnes.fr/COROT/
COSMOS = Cosmic Evolution Survey, http://cosmos.astro.caltech.edu
CRTS = Catalina Real-Time Transients Survey, http://crts.caltech.edu
CSS = Catalina Sky Survey, http://www.lpl.arizona.edu/css
DASCH = Digital Access to a Sky Century at Harvard, http://hea-www.harvard.edu/DASCH
DENIS = Deep Near Infrared Survey of the Southern Sky, http://cdsweb.u-
 strasbg.fr/online/denis.html
DES = Dark Energy Survey, http://www.darkenergysurvey.org
DLS = Deep Lens Survey, http://dls.physics.ucdavis.edu
DSS = Digitized Sky Survey, http://archive.stsci.edu/cgi-bin/dss_form
DPOSS = Palomar Digital Sky Survey, http://www.astro.caltech.edu/~george/dposs
EROS-1,2 = Exp'erience pour la Recherche d'Objets Sombres, http://eros.in2p3.fr/
FGST = Fermi Gamma-ray Space Telescope, http://fermi.gsfc.nasa.gov
FIRST = Faint Images of the Radio Sky at Twenty-Centimeters, http://sundog.stsci.edu
Gaia, http://gaia.esa.int

GALEX = Galaxy Evolution Explorer, http://www.galex.caltech.edu
GAMA = Galaxy And Mass Assembly survey, http://gama-survey.org/
GLIMPSE = Spitzer Galactic Legacy Infrared Mid-Plane Survey Extraordinaire,
  http://www.astro.wisc.edu/sirtf/
GMAN = Global Microlensing Alert Network, http://darkstar.astro.washington.edu/
GMRT = Giant Metrewave Radio Telescope, http://gmrt.ncra.tifr.res.in/
GOODS = The Great Observatories Origins Deep Survey, http://www.stsci.edu/science/goods
GSC = Guide Star Catalog, http://gsss.stsci.edu/Catalogs/Catalogs.htm
HDF = Hubble Deep Field, http://www.stsci.edu/ftp/science/hdf/hdf.html
HEASARC = High Energy Astrophysics Science Archive Research Center,
  http://heasarc.gsfc.nasa.gov/
HIPASS = HI Parkes All Sky Survey, http://aus-vo.org/projects/hipass
HST = Hubble Space Telescope, http://www.stsci.edu/hst
HUDF = Hubble Ultra-Deep Field, http://www.stsci.edu/hst/udf
IceCube = http://icecube.wisc.edu/, a neutrino experiment in Antarctica.
IRAS = Infrared Astronomical Satellite, http://iras.ipac.caltech.edu/IRASdocs/iras.html
IRSA, http://irsa.ipac.caltech.edu/
Kepler, http://kepler.nasa.gov/
LAMOST = Large Sky Area Multi-Object Spectroscopic Telescope, http://www.lamost.org
LINEAR = Lincoln Near Earth Asteroid Research, http://www.ll.mit.edu/mission/space/linear/
LOFAR = LOw Frequency Array, http://www.lofar.org/
LSST = Large Synoptic Survey Telescope, http://www.lsst.org
MACHO = Massive Astrophysical Compact Halo Object, http://wwwmacho.anu.edu.au/
MAST = Multimission Archive at STScI, http://archive.stsci.edu/index.html
MAXI = Monitor of All-sky X-ray Image, http://kibo.jaxa.jp/en/experiment/ef/maxi/
MeerKAT, http://www.ska.ac.za/meerkat/index.php
MicroFun = Microlensing Follow-Up Network, http://www.astronomy.ohio-state.edu/~microfun/
MILAGRO = http://scipp.ucsc.edu/personnel/milagro.html
MOA = Microlensing Observations in Astrophysics, http://www.phys.canterbury.ac.nz/moa/
MOST = Microvariability and Oscillations of Stars, http://www.astro.ubc.ca/MOST/
NDWFS = The NOAO Deep Wide-Field Survey, http://www.noao.edu/noao/noaodeep
NEAT = Near-Earth Asteroid Tracking team, http://neat.jpl.nasa.gov
NED = NASA/IPAC Extragalctic Database, http://ned.ipac.caltech.edu/
NOMAD = Naval Observatory Merged Astrometric Dataset
  http://www.usno.navy.mil/USNO/astrometry/optical-IR-prod/nomad).
NVSS = NRAO VLA Sky Survey, http://www.cv.nrao.edu/nvss
OGLE-I,II,II,IV = Optical Gravitation Lensing Experiment, http://ogle.astrouw.edu.pl/
PANDAS = Pan-Andromeda Archaeological Survey, http://www.nrc-cnrc.gc.ca/eng/projects/hia/pandas.html
PanSTARRS = Panoramic Survey Telescope and Rapid Response System, http://pan-starrs.ifa.hawaii.edu
Pi of the Sky, http://grb.fuw.edu.pl/
PKS = Parkes Radio Sources, http://archive.eso.org/starcat/astrocat/pks.html
PLANET = Probing Lensing Anomalies NETwork, http://www.planet-legacy.org/
PMN = Parkes-MIT-NRAO survey,
  http://www.parkes.atnf.csiro.au/observing/databases/pmn/pmn.html

POSS = Palomar Observatory Sky Survey, http://www.astro.caltech.edu/~wws/poss2.html
PQ = The Palomar-Quest Sky Survey, http://palquest.org
PTF = Palomar Transient Factory, http://www.astro.caltech.edu/ptf
RAPTOR = Rapid Telescopes for Optical Response,
    http://www.lanl.gov/quarterly/q_fall03/observatories.shtml
RASS = ROSAT All Sky Survey, http://heasarc.nasa.gov/docs/rosat/rass.html
RoboNet-I,II, http://www.astro.ljmu.ac.uk/RoboNet/, http://robonet.lcogt.net/
ROTSE = Robotic Optical Transient Search Experiment, http://www.rotse.net
SASSy = SCUBA-2 All-Sky Survey, http://www.jach.hawaii.edu/JCMT/surveys/sassy/
SDSS = Sloan Digital Sky Survey, http://www.sdss.org, http://www.sdss3.org
SKA = Square Kilometer Array, http://www.skatelescope.org
SNFactory = The Nearby Supernova Factory, http://snfactory.lbl.gov/
SpaceWatch, http://spacewatch.lpl.arizona.edu/
SuperWASP = Super Wide Angle Search for Planets, http://www.superwasp.org/
SWIRE = Spitzer Wide-area InfraRed Extragalactic survey, http://swire.ipac.caltech.edu
UKIDSS = UKIRT Infrared Deep Sky Survey, http://www.ukidss.org
USNO = United States Naval Observatory, http://www.usno.nasa.mil/USNO
VISTA = Visible and Infrared Survey Telescope for Astronomy, http://www.vista.ac.uk/
VLA = NRAO Very Large Array, http://www.vla.nrao.edu
VO = Virtual Observatory, http://www.ivoa.net
VLT = Very Large Telescope, http://www.eso.org/public/teles-instr/vlt.html
VST = VLT Survey Telescope, http://www.eso.org/public/teles-instr/surveytelescopes/vst.html
WISE = Wide-Field Infrared Survey Explorer, http://wise.ssl.berkeley.edu
WMAP = Wilkinson Microwave Anisotropy Probe, http://map.gsfc.nasa.gov